\documentclass[%
 reprint,
 amsmath,amssymb,
 aps,
]{revtex4-2}
\usepackage[T1]{fontenc}
\usepackage[utf8]{inputenc}
\usepackage[english]{babel}
\usepackage{bbm}
\usepackage{graphicx}
\usepackage[normalem]{ulem}
\usepackage{tabularx}
\usepackage{multirow}
\usepackage{array}
\newcolumntype{P}[1]{>{\centering\arraybackslash}p{#1}}
\usepackage{longtable}
\usepackage[usenames,dvipsnames]{xcolor}
\usepackage[compat=1.1.0]{tikz-feynman}
\usepackage{hyperref}
\usepackage{feynmp-auto} %draw Feynman graphs
\hypersetup{unicode,psdextra}
\usepackage{scalerel}
\usepackage{mathabx} % \drsh down right arrow for chemical reactions
\usepackage{tikz-feynman} % Draw Feynman diagrams directly written in the file
\usetikzlibrary{decorations.pathreplacing,calligraphy} % Decorations for tikz

\usepackage{natbib}
\newcommand{\msbar}{$\overline{\mathrm{MS}}$~}

\newcommand*{\eweakgroup}{\mbox{$\mathit{SU}(2)_L \times U(1)_Y$}}

\newcommand*{\unitmatrix}{\mathbbm{1}}
\newcommand*{\tvec}[1]{\boldsymbol{#1}}              % 3 vector

\newcommand*{\trans}{\mathrm{T}}                     % transposed

\DeclareMathOperator{\diag}{diag}		%diagonal matrix
\allowdisplaybreaks

\begin{document} 

\title{CMS results for the $\gamma \gamma$ production at the LHC: do they give a hint for a Higgs boson of the maximally CP symmetric two-Higgs-doublet model?}

\author{M.~Maniatis}
\email{maniatis8@gmail.com}
 \affiliation{Centro de Ciencias Exactas,  Universidad del B\'io-B\'io,  Casilla  447,  Chill\'{a}n, Chile}
\author{O.~Nachtmann}
 \email{O.Nachtmann@thphys.uni-heidelberg.de}
\affiliation{Institut f\"ur Theoretische Physik, Universit\"at Heidelberg,
Philosophenweg 16, D-69120 Heidelberg, Germany}

%\author[a]{M.~Maniatis}
%\author[b]{O.~Nachtmann}

%\affiliation[a]{Centro de Ciencias Exactas,  Universidad del B\'io-B\'io,  Casilla  447,  Chill\'{a}n, Chile}
%\affiliation[b]{Institut f\"ur Theoretische Physik, Universit\"at Heidelberg,
%Philosophenweg 16, D-69120 Heidelberg, Germany}
%
%\emailAdd{O.Nachtmann@thphys.uni-heidelberg.de}
%\emailAdd{maniatis8@gmail.com}
 
 \begin{abstract}
Recent measurements of the CMS experiment at the LHC show possibly, with about $3\sigma$ significance, a resonance in di-photon events with an invariant mass of~95.4~GeV.
 If this resonance can be confirmed, this could be a 
hint for a new elementary particle beyond the Standard Model.
An additional Standard-model-like Higgs boson with this mass could be excluded by the CMS experiment. 
We investigate whether this resonance could fit into a two-Higgs-doublet model highly constrained by CP symmetry, the so-called maximally-CP-symmetric model (MCPM). 
In the strict symmetry limit of the MCPM only the fermions of the third generation 
obtain masses. We discuss a mechanism where the first and second generation fermions get masses through 
effects from interactions with very high mass particles. 
The latter can be integrated out at LHC energies giving effective Lagrangian terms, among them these mass terms. 
This procedure also gives us a Cabibbo-Kobayashi-Maskawa (CKM) matrix which automatically satisfies some salient structural features observed in experiments. 
We call the resulting theory~MCPM'. 
Our finding shows that indeed the enhancement measured by CMS could originate from the pseudoscalar Higgs boson $h''$ of this model. 
According to the model the boson $h''$ would mainly be produced in the Drell-Yan reaction by charm-anticharm-quark fusion. 
The main decay mode of $h''$ is predicted to be $h'' \to c\bar{c}$.
We then consider the so called oblique parameters~$S$, $T$, $U$ which give us an allowed region for the mass of the scalar Higgs boson~$h’$ versus that of the charged ones $H^\pm$ of the MCPM'. 
We calculate the effect of these charged bosons $H^\pm$ in the leptonic decays of the
charm mesons $D^\pm$ and the
 charm-strange mesons $D_s^\pm$. Our results indicate that the mass $m_{H^\pm}$ of the charged Higgs bosons $H^\pm$ of the MCPM' should 
 be around~300~GeV.
\end{abstract}
% =============================================================================

\maketitle
\flushbottom

% =============================================================================
\section{Introduction}
The search for physics beyond the Standard Model~(SM) is a central topic of the current experiments at the LHC.
One can, for instance, ask if there exist Higgs bosons in addition to the one Higgs boson of the SM, which by now is well established; see e.g.~\cite{ATLAS:2022vkf}. 
One possible extension of the SM is to a model with two Higgs doublets, a 2HDM. 2HDM's
have a long history and
 got a lot of attention in recent years; see for instance~\cite{Lee:1973iz,Bernreuther:1992dz,Arhrib:1998gr,Gunion:1989we,Krawczyk:2002df,Ferreira:2010hy,Branco:2011iw,Basso:2012st,Krause:2016xku,Basler:2017uxn,Maniatis:2020wfz,Ferreira:2020ana,Wang:2022yhm,Diaz-Cruz:2004wsi,Ferreira:2023dke}. 

An interesting feature of the Higgs sector of 2HDMs is that there one can define so called generalized CP~(GCP) transformations. Due to the presence of two Higgs doublets the standard CP transformation can be augmented by a unitary mixing of the doublets. A convenient classification of GCP transformations has, for instance, been given in~\cite{Maniatis:2007vn}. Such GCP transformations, extended to the full theory including fermions, play an essential role in the construction of the maximally-CP-symmetric model~(MCPM)~\cite{Maniatis:2007de}. This model is the starting point of our present study. 
A short description of this model is given in~\cite{Maniatis:2010sb}.
Phenomenological consequences of the MCPM have been presented by us in~\cite{Maniatis:2009vp, Maniatis:2009by}.
As in all 2HDMs the MCPM contains as physical Higgs bosons three neutral ones, denoted by us as $\rho'$, $h'$, $h''$, and a pair of charged Higgs bosons~$H^\pm$. The boson~$\rho'$ of the MCPM behaves very much like the SM Higgs boson. The scalar $h'$ and the pseudoscalar $h''$ behave quite differently.  Their main production mode in proton--proton collisions is the Drell-Yan reaction $c + \bar{c} \to h', h''$. Important decay modes are $h', h'' \to c \bar{c}$. But there are also the decays $h', h'' \to \gamma \gamma$. Therefore, the reaction $p + p \to \gamma + \gamma + X$ is suitable to search for the neutral Higgs bosons $h'$ and $h''$ of the MCPM. \\

Recently the CMS collaboration has published results for $p + p \to \gamma + \gamma + X$ from the LHC at center-of-mass energy $\sqrt{s} = 13$~TeV~\cite{CMS:2023yay}. The search was for a SM-like Higgs boson in the mass range 70 to 110~GeV. 
A possible enhancement of the $\gamma\gamma$ yield compared to the expectation under the background-only hypothesis has been observed for an invariant $\gamma\gamma$ mass $m_{\gamma\gamma} = 95.4$~GeV; see Figs.~5 and 7 of~\cite{CMS:2023yay}.
As we can see from Fig.~5 of~\cite{CMS:2023yay}
 the possible enhancement of the product cross-section for Higgs production~$\sigma_H$ times branching ratio $\text{B}(H \to \gamma\gamma)$ is of order 
\begin{equation} \label{eq1.1}
\sigma_H \times \text{B}(H \to \gamma\gamma) \approx 0.01- 0.04 \text{ pb.}
\end{equation}
Very interesting are the local $p$ values shown in 
Fig.~7 of~\cite{CMS:2023yay}.
The upward fluctuation 
of $\sigma_H \times \text{B}(H \to \gamma\gamma)$ occurred at the same mass value $m_H = 95.4$~GeV for all CMS runs. 
In~\cite{CMS:2023yay} the local (global) significance of the possible effect observed at 95.4~GeV is given as 2.9~$\sigma$ (1.3~$\sigma$). 
In~\cite{Bhattacharya:2023lmu} a detailed combination of relevant ATLAS, CMS and LEP results was performed and an effect at
approximately~95~GeV
with a global significance of 3.8~$\sigma$ was extracted.
It can be seen from Fig.~5 of~\cite{CMS:2023yay} that an additional SM-like Higgs boson at a mass of 80 to 110 GeV is clearly excluded. 

In the present paper we investigate the question if the possible enhancement of $\gamma\gamma$ production at 95.4~GeV could be due to the production and decay of the boson~$h''$ or $h'$ in the MCPM.
Let us also
mention the discussion of the 2HDM with an additional complex singlet~\cite{Biekotter:2023jld} with respect to the possible resonance at 95.4~GeV. 
In \cite{Coloretti:2023yyq} a model with two Higgs doublets, a real scalar singlet and a Higgs triplet was presented. 
In~\cite{Chen:2023bqr} the Georgi-Machacek model was discussed in connection with the possible effect of beyond SM physics at around~95~GeV.
\\

Our paper is organised as follows. To make the paper self contained we recall in Sec.~\ref{sec:basics} the main features of the MCPM. 
In the strictly symmetric MCPM only the fermions of the third generation get masses unequal to zero. 
In Appendix~\ref{appA} we discuss a mechanism how to introduce nonzero masses for the fermions of the first and second generation as low-energy
remnants of very high mass-scale physics.
In this way we also introduce a non trivial Cabibbo-Kobayashi-Maskawa~(CKM) matrix~$V$.
We call the resulting theory~MCPM'. Throughout our work we neglect neutrino masses and mixings. 
Sec.~\ref{sec:results} presents our results for an MCPM' Higgs boson 
of 95.4~GeV mass.
 In Sec.~\ref{sec:comp} we compare with experiment and discuss our findings. 
Sec.~\ref{sec:Ds} deals with leptonic decays of the
charm mesons~$D^\pm$ and the
 charm-strange mesons~$D_s^\pm$ in view of effects of the charged bosons~$H^\pm$ of the MCPM' in these decays.
Our conclusions are drawn in Sec.~\ref{sec:con}. 

%%%%%%%%%%%%%%%%%%%%%%%%%%%%%%%%%%%%%%%%%%%%%%%%
%%%%%%%%%%%%%%%%%%%%%%%%%%%%%%%%%%%%%%%%%%%%%%%%
\section{Basics of the MCPM}
\label{sec:basics}

The MCPM has been introduced first in~\cite{Maniatis:2007de} and a short description of it can be found in~\cite{Maniatis:2010sb}. 
For the convenience of the reader, we outline here the essential features of the MCPM. 

The MCPM is a two-Higgs-doublet model, 2HDM. It has the two Higgs-boson doublets,
\begin{equation} \label{eq2.1}
\varphi_1 = \begin{pmatrix} \varphi_1^+\\ \varphi_1^0 \end{pmatrix}, \quad
\varphi_2 = \begin{pmatrix} \varphi_2^+\\ \varphi_2^0 \end{pmatrix}.
\end{equation}
Both are assumed to have weak hypercharge $y= 1/2$. It is convenient to use the $K$ formalism, which works with the gauge-invariant bilinears~\cite{Nagel:2004sw, Maniatis:2006fs, Maniatis:2007vn},
\begin{gather} \label{eq2.2}
\begin{split}
K_0 = \varphi_1^\dagger \varphi_1 + \varphi_2^\dagger \varphi_2 =
\begin{pmatrix} \varphi_1^\dagger, \varphi_2^\dagger \end{pmatrix} 
\begin{pmatrix} \varphi_1\\ \varphi_2 \end{pmatrix}, \\
\tvec{K} = \begin{pmatrix} K_1\\ K_2\\ K_3 \end{pmatrix} =
\begin{pmatrix} 
\varphi_1^\dagger \varphi_2 + \varphi_2^\dagger \varphi_1\\
i \varphi_2^\dagger \varphi_1 - i \varphi_1^\dagger \varphi_2\\
\varphi_1^\dagger \varphi_1 - \varphi_2^\dagger \varphi_2
\end{pmatrix}
=
\begin{pmatrix} \varphi_1^\dagger, \varphi_2^\dagger \end{pmatrix}  \tvec{\sigma}
\begin{pmatrix} \varphi_1\\ \varphi_2 \end{pmatrix}.
\end{split}
\end{gather}
Here $\tvec{\sigma}=(\sigma^1, \sigma^2, \sigma^3)^\trans$ and $\sigma^a$ $(a=1,2,3)$ are 
the Pauli matrices.
Basis changes of the Higgs fields~\eqref{eq2.1}
\begin{equation} \label{eq2.2a}
\varphi_i(x) \to \varphi_i'(x) = U_{ij} \varphi_j(x), \quad
U = \left(U_{ij}\right) \in U(2),
\end{equation}
correspond to $SO(3)$ rotations in $K$ space 
\begin{equation} \label{eq2.2b}
\begin{split}
&K_0(x) \to K_0'(x) = K_0(x),\\
&K_a(x) \to K_a'(x) = R_{ab}(U) K_b(x), 
\end{split}
\end{equation}
where
\begin{equation} \label{eq2.2c}
U^\dagger \sigma^a U = R_{ab}(U) \sigma^b, \quad
R(U) = \left( R_{ab}(U) \right) \in SO(3).
\end{equation}
In this formalism the most general Higgs potential reads
\begin{equation} \label{eq2.3}
V = \xi_0 K_0 + \tvec{\xi}^\trans \tvec{K} + \eta_{00} K_0^2 
+ 2 K_0 \tvec{\eta}^\trans \tvec{K} + \tvec{K}^\trans E \tvec{K}.
\end{equation}
Here the parameters $\xi_0$, $\eta_{00}$, the 3-component vectors
$\tvec{\xi}$, $\tvec{\eta}$, and the symmetric $3 \times 3$ matrix $E = E^\trans$ are all real. 

Now we come to generalised CP (GCP) transformations which are generically of the form
\begin{multline} \label{eq2.4}
\varphi_i(x) \to U_{ij} \varphi_j^*(x'), \quad
x = (x^0, \tvec{x})^\trans, \quad
x' = (x^0, -\tvec{x})^\trans, 
\\
U = \left(U_{ij}\right) \in U(2). 
\end{multline}
As can be seen from~\eqref{eq2.2} in $K$ space this corresponds to
\begin{equation} \label{eq2.5}
K_0(x) \to K_0(x'), \quad
\tvec{K}(x) \to \tilde{R} \tvec{K}(x'),
\end{equation}
where $\tilde{R}$ is an improper rotation matrix, $\det(\tilde{R}) = -1$. 
For a GCP transformation we require that applying it twice gives back the unit transformation in $K$ space, $\tilde{R}^2 = \unitmatrix_3$. 
This leads to two types of GCPs:
\begin{equation} \label{eq2.6}
\begin{split}
(i) \qquad  \tilde{R} =& -\unitmatrix_3, \text{ point reflection in } K \text{ space,}\\
(ii) \qquad  \tilde{R} =& R^\trans \tilde{R}_2 R, \text{ reflection on a plane in } K \text{ space,} \\
& \text{ where } \tilde{R}_2 = \diag(1, -1, 1), \quad R \in SO(3)\;.
\end{split}
\end{equation}

We call GCP transformations~\eqref{eq2.4} $\text{CP}_g^{(i)}$ and 
$\text{CP}_g^{(ii)}$ if they satisfy~$(i)$ and $(ii)$, respectively. 
The standard CP transformation for the Higgs fields has $U_{ij} = \delta_{ij}$ in \eqref{eq2.4}
corresponding to $\tilde{R} = \tilde{R}_2$ in \eqref{eq2.5}.
It is, thus, a $\text{CP}_g^{(ii)}$ transformation since, according to~\eqref{eq2.6}, it induces 
a reflection on the 1—3 plane in $K$ space.

Setting $U_{ij} = \epsilon_{ij}$ in~\eqref{eq2.4}, where
\begin{equation} \label{eq10a}
\left( \epsilon_{ij} \right) = 
\begin{pmatrix} \phantom{+}0 & 1\\ -1 & 0 \end{pmatrix}
\end{equation}
we get the $\text{CP}_g^{(i)}$ transformation which gives 
$\tilde{R} = - \unitmatrix_3$.

In~\cite{Maniatis:2007de} the question was studied if one could have a 2HDM
 allowing the GCP of type~$(i)$, $\text{CP}_g^{(i)}$, as an exact symmetry and how the corresponding symmetric Yukawa
 term in the Lagrangian would look like. 
 
 The potential of such a theory, which was called the {\em maximally-CP-symmetric model}~(MCPM), must be invariant under $\tvec{K} \to -\tvec{K}$ and, therefore, see \eqref{eq2.3}, 
 \begin{equation} \label{eq2.7}
 V_{\text{MCPM}} = \xi_0 K_0 + \eta_{00} K_0^2 + \tvec{K}^\trans E \tvec{K}.
 \end{equation}
 We can choose a basis for the Higgs fields, where $E = E^\trans$ is diagonal,
 \begin{equation} \label{eq2.8}
 E = \diag(\mu_1, \mu_2, \mu_3), 
\end{equation}
and the eigenvalues are ordered 
\begin{equation} \label{eq2.9}
\mu_1 \ge \mu_2 \ge \mu_3.
\end{equation}
%\begin{equation}
% \begin{split}
% V_{\text{MCPM}}   & =
% m^2 (\varphi_1^\dagger \varphi_1 + \varphi_2^\dagger \varphi_2)
% + \frac{1}{2} \lambda_1 \left(  
% (\varphi_1^\dagger \varphi_1)^2 +  (\varphi_2^\dagger \varphi_2)^2 \right) 
% \\ & +
% \lambda_3  (\varphi_1^\dagger \varphi_1)  (\varphi_2^\dagger \varphi_2)
%+
% \lambda_4  (\varphi_1^\dagger \varphi_2)  (\varphi_2^\dagger \varphi_1)
% \\ & +
%\frac{1}{2}
% \lambda_5  \left( 
% (\varphi_1^\dagger \varphi_2)^2 +  (\varphi_2^\dagger \varphi_1)^2 \right).
% \end{split}
% \end{equation}
% with
% \begin{multline}
% \xi_0 = m^2, \quad
% \eta_{00} = \frac{1}{4} (\lambda_1 + \lambda_3),\\
% E = \frac{1}{4} \diag ( \lambda_4 + \lambda_5, \lambda_4 - \lambda_5, \lambda_1 - \lambda_3),
% \end{multline}
% where all parameters, $m^2$, $\lambda_1$, $\lambda_3$, $\lambda_4$, $\lambda_5$ 
% are real.

From theorem~5 of \cite{Maniatis:2007vn} we know that the potential~\eqref{eq2.7}
leads to a stable theory with the correct electroweak symmetry breaking~(EWSB) and no zero mass  charged Higgs boson if and only if
\begin{multline} \label{eq2.10}
\eta_{00} >0, \quad
\mu_a + \eta_{00} >0, \text{ for } a = 1,2,3, \\
\xi_0 <0, \quad
\mu_3 < 0\;.
\end{multline}
The $\text{CP}_g^{(i)}$ symmetry is automatically spontaneously broken by EWSB. 
In~\cite{Maniatis:2007de} the couplings of the Higgs fields~\eqref{eq2.1} to fermions were studied, requiring invariance under the $\text{CP}_g^{(i)}$ symmetry transformation. 
It turned out that for a single fermion family this symmetry requires zero coupling. 
This finding can be interpreted as giving us a symmetry reason for the existence of more than one fermion family in nature. 

In~\cite{Maniatis:2007de} strict inequalities were required in~\eqref{eq2.10} which assured absence of zero mass Higgs bosons and of mass degeneracy of $h'$ and $h''$. We see then that the potential~\eqref{eq2.7} is not only~$\text{CP}_g^{(i)}$ symmetric but has in addition exactly three invariances~$\text{CP}_{g\; j}^{(ii)}$ ($j=1,2,3$) corresponding to the reflections on the coordinate planes in~$\tvec{K}$ space. We write these transformations for the Higgs fields~$\varphi_i(x)$ \eqref{eq2.1} and the bilinears $K_0(x)$, $\tvec{K}(x)$ generically as follows 
(see \eqref{eq2.4}, \eqref{eq2.5}):
\begin{equation} \label{16a}
\begin{split}
\text{CP}_g:  \qquad 
& \varphi_i(x) \to W_{ij} \varphi_j^*(x'),\\
& K_0(x) \to K_0(x'),\\
& \tvec{K}(x) \to \tilde{R} \tvec{K}(x').
\end{split}
\end{equation}
In Table~\ref{tabW} we give the matrices $W=(W_{ij})$ and $\tilde{R}$ for the above $\text{CP}_g$ transformations (see table 1 of~\cite{Maniatis:2007de})
\begin{table}
\begin{tabular}{ccc}
$\text{CP}_g$ & $W$ & $\tilde{R}$ \\
\hline
$\text{CP}_g^{(1)}$ \quad& $\epsilon$ \quad & $-\unitmatrix_3$\\
$\text{CP}_{g 1}^{(ii)}$ & $\sigma^3$ & $\diag(-1,1,1)$\\
$\text{CP}_{g 2}^{(ii)}$ & $\unitmatrix_2$ & $\diag(1,-1,1)$\\
$\text{CP}_{g 3}^{(ii)}$ & $\sigma^1$ & $\diag(1,1,-1)$\\
\end{tabular}
\caption{\label{tabW} The matrices $W$ and $\tilde{R}$ from~\eqref{16a} for the four $\text{CP}_g$ symmetries of the Higgs potential~\eqref{eq2.7}.}
\end{table}
In~\cite{Maniatis:2007de} the question was asked if one could extend the four $\text{CP}_g$ symmetries of Table~\ref{tabW} to a full theory with two fermion families. Requiring in addition absence of large flavor-changing neutral currents and of mass degeneracy of corresponding fermions of the two families a unique solution was found. One
family could have non-zero masses, the other one had  to be massless. Adding a third family, uncoupled to the Higgs fields, the MCPM was obtained. For the details of these argumentations we refer to~\cite{Maniatis:2007de}.

Before EWSB, the Yukawa couplings of the MCPM are highly symmetric. The third family, $t$, $b$, $\tau$, couples to the Higgs field~$\varphi_1$ proportional to the masses $m_t$, $m_b$, $m_\tau$, respectively.
The second family, $c$, $s$, $\mu$ is massless in the strict symmetry limit and couples to the Higgs field~$\varphi_2$ but proportional to the masses of the {\em third} family, $m_t$, $m_b$, $m_\tau$. See~\eqref{A25} of Appendix~\ref{appA}. 
The first family $u$, $d$, $e$ is uncoupled to the Higgs fields 
and is also massless 
in the strict symmetry limit.

The above identification of the families observed in nature with the families of the MCPM is part of the basic assumptions of this model. As always, the assumptions made in the construction of a model must get their justification from the comparison of theory with experiment.

The Cabibbo-Kobayashi-Maskawa (CKM) Matrix $V$, 
\begin{equation} \label{14.1}
V = \left(V_{ij}\right) = \begin{pmatrix} 
V_{ud} & V_{us} & V_{ub}\\
V_{cd} & V_{cs} & V_{cb}\\
V_{td} & V_{ts} & V_{tb}
\end{pmatrix},
\end{equation}
is required to be the unit matrix in the strict symmetry limit of the MCMP, 
\begin{equation} \label{14.2}
V_{\text{MCPM}} = \unitmatrix_3\;.
\end{equation}
\begin{center}
\begin{table}
\begin{tabular}{lll}
\hline
$m_e = 5.110 \cdot 10^{-4}$&
$m_u = 2.16 \cdot 10^{-3}$&
$m_d = 4.67 \cdot 10^{-3}$\\
$m_\mu = 1.057 \cdot 10^{-1}$&
$m_c = 1.27$&
$m_s = 9.34 \cdot 10^{-2}$\\
$m_\tau = 1.777$&
$m_t = 172.69$&
$m_b = 4.18$\\
\hline
\end{tabular}
\caption{\label{tablemass}
Central values for fermion masses in~GeV as quoted by PDG~\cite{ParticleDataGroup:2024cfk}. For leptons we have rounded to four significant figures.}
\end{table}
\end{center}

Clearly, the strict symmetry limit of the MCPM cannot give an exact representation of what is observed in nature. 
But, as discussed in~\cite{Maniatis:2007de}, the MCPM could be a first approximation to what is observed. 
Indeed, the masses of the first- and second-family fermions are rather small compared to those of the corresponding third-family masses. Using the central mass values as quoted in PDG~\cite{ParticleDataGroup:2024cfk}, see Table~\ref{tablemass}, 
we get:
\begin{equation} \label{14.3}
\begin{split}
&\frac{m_e}{m_\tau} = 2.88 \times 10^{-4}, \quad
\frac{m_\mu}{m_\tau} = 5.95 \times 10^{-2}, \\
& \frac{m_u}{m_t} = 1.25 \times 10^{-5}, \quad
 \frac{m_c}{m_t} = 7.35 \times 10^{-3}, \\
& \frac{m_d}{m_b} = 1.12 \times 10^{-3}, \quad
\frac{m_s}{m_b} = 2.23 \times 10^{-2}\;.
\end{split}
\end{equation}
Note that here the quark masses, except for the top-quark, are \msbar masses.
Also, the CKM matrix~$V$ is close to the unit matrix. This is best seen in the 
Wolfenstein parametrisation~\cite{Wolfenstein:1983yz,Buras:1994ec,Charles:2004jd,ParticleDataGroup:2024cfk}
where
\begin{equation} \label{14.4}
V = \begin{pmatrix} 
1-\lambda^2/2 & \lambda & A \lambda^3(\rho - i \eta)\\
- \lambda & 1-\lambda^2/2 & A \lambda^2\\
A \lambda^3(1-\rho - i \eta) & - A \lambda^2 & 1
\end{pmatrix} + {\cal O}(\lambda^4)
\end{equation}
with
\begin{equation} \label{14.5}
\lambda \approx 0.23, \quad A \approx 0.83, \quad \rho \approx 0.16, 
\quad \eta \approx 0.35;
\end{equation}
see (12.5) and (12.26) in the review~12 of~\cite{ParticleDataGroup:2024cfk}. 

In~\cite{Maniatis:2009vp,Maniatis:2009by,Maniatis:2010sb} phenomenological predictions, respectively estimates, for the properties of the physical Higgs bosons of the MCPM were made.
Where necessary for phase space reasons fermion masses were introduced by hand. But we think that,
nevertheless, this procedure gave us reasonable estimates.

In our present article, on the other hand, we discuss in Appendix~\ref{appA} a way how we can generate non-zero masses in the MCPM also for the first and second generations of fermions. We introduce a breaking of the 
$\text{CP}_g^{(i)}$ 
 symmetry at a very high mass scale. 
 Then the masses of the first and second generation are obtained as remnants of the high-mass 
 physics which explains their smallness relative to the third generation masses; see~\eqref{14.3}. 
 We also generate a non trivial CKM matrix~$V$ in this way.
 In this scenario we can furthermore understand EWSB  and the closely connected spontaneous $\text{CP}_g^{(i)}$ 
 violation as such a remnant of the high-mass physics. 
 The resulting theory for LHC energies is then like the MCPM for the Higgs-boson couplings among themselves and to the $Z$ and $W$ bosons. For the Higgs-boson fermion couplings the {\em large} couplings are as in the MCPM. That is, the boson $\varphi_1$ couples mainly to the third generation of fermions, the boson~$\varphi_2$ mainly to the second generation of fermions. 
 Small couplings of these bosons to other generations are now introduced in a consistent way.  
 We call the resulting theory the MCPM'.
  
 In the following we shall mainly be interested in reactions dominated by the {\em large} couplings as defined above. 
We want to analyse what this MCPM' can say concerning the $\gamma\gamma$ events reported in~\cite{CMS:2023yay}.  
We shall then discuss the leptonic decays of the charm mesons~$D^\pm$ and the charm-strange mesons~$D_s^\pm$ in the MCPM'. 
We also give theoretical estimates for observables where the MCPM' Higgs bosons should show up. 

As already mentioned in the introduction we have in the MCPM' after EWSB five physical Higgs bosons, three neutral ones $\rho'$, $h'$, $h''$, and the charged pair $H^\pm$. With the standard Higgs-field vacuum-expectation value
\begin{equation} \label{14a}
v_0 = \sqrt{\frac{-\xi_0}{\eta_{00}+\mu_3}} = 246 \text{ GeV}
\end{equation}
 we find for the Higgs masses
\begin{equation} \label{eq2.11}
\begin{split}
&m_{\rho'}^2 = 2 (- \xi_0) = 2 v_0^2 (\eta_{00} + \mu_3), \\
&m_{h'}^2 = 2 v_0^2 (\mu_1 -\mu_3), \\
&m_{h''}^2 = 2 v_0^2 (\mu_2 -\mu_3), \\
&m_{H^\pm}^2 = 2 v_0^2 (- \mu_3)\;.
\end{split}
\end{equation}
A strict prediction of the MCPM' is the following mass relation between the pseudoscalar $h''$ and the scalar $h'$,
\begin{equation} \label{eq2.12}
m_{h''} \le m_{h'}\;.
\end{equation}
The Feynman rules for these bosons 
in the MCPM
are given explicitly in Appendix~A of~\cite{Maniatis:2009vp}. 

In the MCPM' these rules stay the same except for the boson-fermion couplings which we can read off from~\eqref{A48}--\eqref{A52} and \eqref{A77} in Appendix~\ref{appA}.
The main feature of interest to us here is that $h'$ and $h''$ have a scalar, respectively, pseudoscalar coupling to charm quarks with coupling constants proportional to the $t$ quark mass divided by~$v_0$; see Fig~\ref{fig3}. Both, $h'$ and $h''$, have the largest coupling to the charm quark. Numerically we have
\begin{equation} \label{2.16}
c'_c = c''_c = \frac{m_t}{v_0} = \frac{173~\text{GeV}}{246~\text{GeV}} = 0.703 \;.
\end{equation}
This is of the order of an electromagnetic coupling
\begin{equation} \label{2.17}
e = \sqrt {4 \pi \alpha_{\text{em}}} = 0.303\;.
\end{equation}

\begin{figure}[!ht]
\centering
\begin{tabularx}{0.8\columnwidth}{p{0.3\columnwidth} p{0.1\columnwidth} p{0.3\columnwidth} p{0.1\columnwidth} }
\raisebox{-.45\height}{
\begin{tikzpicture}
  \begin{feynman}
  %place first vertices
    \vertex (a) at (0,0);
     \vertex (i1) at (-1, 1) ;
     \vertex (i2) at (-1, -1) ;
 	 \vertex (f) at (1, 0){\(h'\)};
   \diagram* {
      (i1) -- [anti fermion, thick, edge label=\(f\)] (a),
      (a) -- [anti fermion, thick, edge label=\(f\)] (i2),
      (a) -- [scalar, thick] (f),
 };
   \end{feynman}
\end{tikzpicture}
}
&
$i c_f'$
&
\raisebox{-.45\height}{
\begin{tikzpicture}
  \begin{feynman}
  %place first vertices
    \vertex (a) at (0,0);
     \vertex (i1) at (-1, 1) ;
     \vertex (i2) at (-1, -1) ;
 	 \vertex (f) at (1, 0){\(h''\)};
   \diagram* {
      (i1) -- [anti fermion, thick, edge label=\(f\)] (a),
      (a) -- [anti fermion, thick, edge label=\(f\)] (i2),
      (a) -- [scalar, thick] (f),
 };
   \end{feynman}
\end{tikzpicture}
}
&
$c_f'' \gamma_5$
\end{tabularx}
\caption{\label{fig3}
Couplings of the second-family fermions $f = \mu, c, s$ to the bosons $h'$ and $h''$. 
We have $c'_\mu = m_\tau/v_0$, $c'_c = m_t/v_0$, $c'_s = m_b/v_0$ and
$c''_\mu = - m_\tau/v_0$, 
$c''_c = m_t/v_0$,
$c''_s = - m_b/v_0$. Here $v_0= 246$~GeV is the standard Higgs vacuum-expectation value.
}
\end{figure}

\renewcommand{\tablename}{Figure} %Set tablename to figure temporarily
\renewcommand{\thetable}{2} % Also use numbers as in Figures temporarily
%%%%%%%%%%%%%%%%%%
\setlength{\LTcapwidth}{\columnwidth} %shorten caption in longtable environment
\begin{longtable}{ll}
\caption{Couplings of $H^\pm$ to fermions in the MCPM'. 
The arrows on the $H$ lines indicate the flow direction of $H^-$.} 
\label{fig100}\\
\endlastfoot
\raisebox{-.45\height}{
\begin{tikzpicture}
  \begin{feynman}
  %place first vertices
    \vertex (a) at (0,0);
     \vertex (i1) at (-1, 1) ;
     \vertex (i2) at (-1, -1) ;
 	 \vertex (f) at (1, 0){\(H\)};
   \diagram* {
      (i1) -- [anti fermion, thick, edge label=\(\nu_\mu\)] (a),
      (a) -- [anti fermion, thick, edge label=\(\mu\)] (i2),
      (a) -- [charged scalar, thick] (f),
 };
   \end{feynman}
\end{tikzpicture}
}
&
$i \frac{m_\tau}{\sqrt{2} v_0} (1+\gamma_5)$
\\ \\
\raisebox{-.45\height}{
\begin{tikzpicture}
  \begin{feynman}
  %place first vertices
    \vertex (a) at (0,0);
     \vertex (i1) at (-1, 1) ;
     \vertex (i2) at (-1, -1) ;
 	 \vertex (f) at (1, 0){\(H\)};
   \diagram* {
      (i1) -- [anti fermion, thick, edge label=\(\mu\)] (a),
      (a) -- [anti fermion, thick, edge label=\(\nu_\mu\)] (i2),
      (f) -- [charged scalar, thick] (a),
 };
   \end{feynman}
\end{tikzpicture}
}
&
$i \frac{m_\tau}{\sqrt{2} v_0} (1-\gamma_5)$
\\ \\
\raisebox{-.45\height}{
\begin{tikzpicture}
  \begin{feynman}
  %place first vertices
    \vertex (a) at (0,0);
     \vertex (i1) at (-1, 1) ;
     \vertex (i2) at (-1, -1) ;
 	 \vertex (f) at (1, 0){\(H\)};
   \diagram* {
      (i1) -- [anti fermion, thick, edge label=\(c\)] (a),
      (a) -- [anti fermion, thick, edge label=\(s\)] (i2),
      (a) -- [charged scalar, thick] (f),
 };
   \end{feynman}
\end{tikzpicture}
}
&
$-i \frac{V_{22}}{\sqrt{2} v_0} \left[m_t (1-\gamma_5) - m_b (1+\gamma_5)\right]$
\\ \\
\raisebox{-.45\height}{
\begin{tikzpicture}
  \begin{feynman}
  %place first vertices
    \vertex (a) at (0,0);
     \vertex (i1) at (-1, 1) ;
     \vertex (i2) at (-1, -1) ;
 	 \vertex (f) at (1, 0){\(H\)};
   \diagram* {
      (i1) -- [anti fermion, thick, edge label=\(s\)] (a),
      (a) -- [anti fermion, thick, edge label=\(c\)] (i2),
      (a) -- [anti charged scalar, thick] (f),
 };
   \end{feynman}
\end{tikzpicture}
}
&
$-i \frac{V^{*}_{22}}{\sqrt{2} v_0} \left[m_t (1+\gamma_5) - m_b (1-\gamma_5)\right]$
\\ \\
\raisebox{-.45\height}{
\begin{tikzpicture}
  \begin{feynman}
  %place first vertices
    \vertex (a) at (0,0);
     \vertex (i1) at (-1, 1) ;
     \vertex (i2) at (-1, -1) ;
 	 \vertex (f) at (1, 0){\(H\)};
   \diagram* {
      (i1) -- [anti fermion, thick, edge label=\(c\)] (a),
      (a) -- [anti fermion, thick, edge label=\(d\)] (i2),
      (f) -- [anti charged scalar, thick] (a),
 };
   \end{feynman}
\end{tikzpicture}
}
&
$-i \frac{V_{21}}{\sqrt{2} v_0} m_t (1-\gamma_5)$
\\ \\
\raisebox{-.45\height}{
\begin{tikzpicture}
  \begin{feynman}
  %place first vertices
    \vertex (a) at (0,0);
     \vertex (i1) at (-1, 1) ;
     \vertex (i2) at (-1, -1) ;
 	 \vertex (f) at (1, 0){\(H\)};
   \diagram* {
      (i1) -- [anti fermion, thick, edge label=\(d\)] (a),
      (a) -- [anti fermion, thick, edge label=\(c\)] (i2),
      (f) -- [charged scalar, thick] (a),
 };
   \end{feynman}
\end{tikzpicture}
}
&
$-i \frac{V_{21}^*}{\sqrt{2} v_0}  m_t (1+\gamma_5)$
\\ \\
\raisebox{-.45\height}{
\begin{tikzpicture}
  \begin{feynman}
  %place first vertices
    \vertex (a) at (0,0);
     \vertex (i1) at (-1, 1) ;
     \vertex (i2) at (-1, -1) ;
 	 \vertex (f) at (1, 0){\(H\)};
   \diagram* {
      (i1) -- [anti fermion, thick, edge label=\(c\)] (a),
      (a) -- [anti fermion, thick, edge label=\(b\)] (i2),
      (f) -- [anti charged scalar, thick] (a),
 };
   \end{feynman}
\end{tikzpicture}
}
&
$-i \frac{V_{23}}{\sqrt{2} v_0}  m_t (1-\gamma_5)$
\\ \\
\raisebox{-.45\height}{
\begin{tikzpicture}
  \begin{feynman}
  %place first vertices
    \vertex (a) at (0,0);
     \vertex (i1) at (-1, 1) ;
     \vertex (i2) at (-1, -1) ;
 	 \vertex (f) at (1, 0){\(H\)};
   \diagram* {
      (i1) -- [anti fermion, thick, edge label=\(b\)] (a),
      (a) -- [anti fermion, thick, edge label=\(c\)] (i2),
      (f) -- [charged scalar, thick] (a),
 };
   \end{feynman}
\end{tikzpicture}
}
&
$-i \frac{V_{23}^*}{\sqrt{2} v_0}  m_t (1+\gamma_5)$
\\ \\
\raisebox{-.45\height}{
\begin{tikzpicture}
  \begin{feynman}
  %place first vertices
    \vertex (a) at (0,0);
     \vertex (i1) at (-1, 1) ;
     \vertex (i2) at (-1, -1) ;
 	 \vertex (f) at (1, 0){\(H\)};
   \diagram* {
      (i1) -- [anti fermion, thick, edge label=\(u\)] (a),
      (a) -- [anti fermion, thick, edge label=\(s\)] (i2),
      (f) -- [anti charged scalar, thick] (a),
 };
   \end{feynman}
\end{tikzpicture}
}
&
$i \frac{V_{12}}{\sqrt{2} v_0}  m_b (1+\gamma_5)$
\\ \\
\raisebox{-.45\height}{
\begin{tikzpicture}
  \begin{feynman}
  %place first vertices
    \vertex (a) at (0,0);
     \vertex (i1) at (-1, 1) ;
     \vertex (i2) at (-1, -1) ;
 	 \vertex (f) at (1, 0){\(H\)};
   \diagram* {
      (i1) -- [anti fermion, thick, edge label=\(s\)] (a),
      (a) -- [anti fermion, thick, edge label=\(u\)] (i2),
      (f) -- [charged scalar, thick] (a),
 };
   \end{feynman}
\end{tikzpicture}
}
&
$i \frac{V_{12}^*}{\sqrt{2} v_0}  m_b (1-\gamma_5)$
\\ \\
\raisebox{-.45\height}{
\begin{tikzpicture}
  \begin{feynman}
  %place first vertices
    \vertex (a) at (0,0);
     \vertex (i1) at (-1, 1) ;
     \vertex (i2) at (-1, -1) ;
 	 \vertex (f) at (1, 0){\(H\)};
   \diagram* {
      (i1) -- [anti fermion, thick, edge label=\(t\)] (a),
      (a) -- [anti fermion, thick, edge label=\(s\)] (i2),
      (f) -- [anti charged scalar, thick] (a),
 };
   \end{feynman}
\end{tikzpicture}
}
&
$i \frac{V_{32}}{\sqrt{2} v_0}  m_b (1+\gamma_5)$
\\ \\
\raisebox{-.45\height}{
\begin{tikzpicture}
  \begin{feynman}
  %place first vertices
    \vertex (a) at (0,0);
     \vertex (i1) at (-1, 1) ;
     \vertex (i2) at (-1, -1) ;
 	 \vertex (f) at (1, 0){\(H\)};
   \diagram* {
      (i1) -- [anti fermion, thick, edge label=\(s\)] (a),
      (a) -- [anti fermion, thick, edge label=\(t\)] (i2),
      (f) -- [charged scalar, thick] (a),
 };
   \end{feynman}
\end{tikzpicture}
}
&
$i \frac{V_{32}^*}{\sqrt{2} v_0}  m_b (1-\gamma_5)$
\end{longtable}
\renewcommand{\tablename}{Table} %Set back tablename to Table
\renewcommand{\thetable}{\Roman{table}} % Also set back to latin.
\addtocounter{figure}{1} %Increase figure counter
\addtocounter{table}{-1} %Decrease table counter

Of course, parity is not conserved in the MCPM' and the naming of $h'$ $(h'')$ scalar (pseudoscalar) is only a reminder of their scalar (pseudoscalar) coupling to fermions; see Fig.~\ref{fig3}.

For the discussion of the leptonic decays of the mesons~$D^\pm$ and $D_s^\pm$ in the MCPM' we also need the couplings of $H^\pm$ to fermions which we give in Fig.~\ref{fig100}; see~\eqref{A77} of the Appendix~\ref{appA}. 
%appendix~A of~\cite{Maniatis:2009vp}.
In the MCPM $H^\pm$ only couple, concerning the fermions, to the $\mu \nu_\mu$ and $c s$ combinations. In the MCPM' these couplings are the dominant ones. Relative to these the couplings to other fermions are 
suppressed by off-diagonal matrix elements of the CKM matrix; see~\eqref{A77}.
%expected to be at a level of the small mass ratios of~\eqref{14.3}. 

%%%%%%%%%%%%%%%%%%%%%%%%%%%%%%%%%%%%%%
%%%%%%%%%%%%%%%%%%%%%%%%%%%%%%%%%%%%%%
\section{Results}
\label{sec:results} 
Since a mass value of 95.4~GeV seems rather low for a new Higgs boson we shall investigate in this chapter the consequences of having in the framework of the MCPM' the boson $h''$ at this mass. 
The boson $h'$ must then have a higher or equal mass; see~\eqref{eq2.12}.
According to Fig.~8 of~\cite{Maniatis:2009vp} 
the main decay mode of~$h''$ with a mass around 100~GeV is $h'' \to c \bar{c}$. All other decay modes have a branching fraction below~$10^{-3}$. 
Thus we have for
 the expected total width $\Gamma_{h''}$
of such a $h''$, 
with an accuracy at the per mille level,
$\Gamma_{h''} \approx \Gamma(h'' \to c \bar{c})$ and,
therefore, from Table~3 of~\cite{Maniatis:2009vp}, with $m_{h''}$ in units of GeV,
\begin{equation} \label{eq3.1}
\begin{split}
\Gamma_{h''} = &12.08 \times \frac{m_{h''}}{200} \text{ GeV }\\
= &
5.76 \text{ GeV for } m_{h''} = 95.4 \text{ GeV.} 
\end{split}
\end{equation}
Its $\gamma\gamma$ width has been calculated in Sec.~3.3, (3.9)—(3.17),
 of~\cite{Maniatis:2009vp}. 
For 95.4~GeV we get for this width and the corresponding branching fraction
\begin{equation} \label{eq3.2}
\Gamma(h'' \to \gamma\gamma) =  3.26~\text{keV},\quad
\text{B}(h'' \to \gamma\gamma) = 5.66 \cdot 10^{-7}\;.
\end{equation}

Now we investigate the production of $h''$ in $pp$ collisions followed by the decay $h'' \to \gamma\gamma$,
\begin{equation} \label{eq3.3}
\begin{split}
p(p_1, s_1) + p(p_2, s_2) \boldsymbol{\to}  &h''(k) + X\;,\\
& \;\scaleto{\drsh}{8pt} \gamma(k_1, \epsilon_1) + \gamma(k_2, \epsilon_2)
\\ k = k_1 + k_2\;.
\end{split}
\end{equation}
Here $p_1$, $p_2$, $k$, $k_1$, and $k_2$ are the momenta of the particles, $\epsilon_1$ and $\epsilon_2$ are the
polarisation vectors of the photons, and $s_1$, $s_2$ are the 
spin indices
of the protons. 

We are interested in the cross section with respect to the invariant $\gamma\gamma$ mass squared
\begin{equation} \label{eq3.5}
m_{\gamma\gamma}^2 = (k_1 + k_2)^2\;.
\end{equation}
We assume unpolarised initial protons, no observation of photon polarisations in the final state, and
use $\Gamma_{h''} \ll m_{h''}$; see~\eqref{eq3.1}. 
Furthermore, we assume measurement of $h''$, that is, 
the $\gamma\gamma$ system, in a certain phase-space region ${\cal B}$. We define the inclusive cross 
section for $h''$ production by
\begin{multline} \label{eq3.6}
d \sigma_{\text{inc}} \left( p(p_1) + p(p_2) \to h''(k) + X \right)
=
\frac{d^3 k}{k^0} f_{\text{inc}}(k)
=\\ 
\frac{d^3 k}{k^0} \frac{1}{2 \sqrt{s(s- 4 m_p^2)}} \frac{1}{2} \frac{1}{(2\pi)^3} 
\sum_X (2\pi)^4 \delta^{(4)} (k + p_X - p_1 -p_2) 
\\ 
\times \frac{1}{4} \sum_{\text{spins}}
\left| \langle h''(k), X(p_X) | T | p(p_1, s_1), p(p_2, s_2) \rangle \right|^2\;,
\end{multline}
where $s=(p_1+p_2)^2$ is the center-of-mass energy squared. The $\gamma\gamma$ width of $h''$ is given by
\begin{multline} \label{eq3.7}
\Gamma(h'' \to \gamma\gamma) =
\frac{1}{ 2 m_{h''}} \frac{1}{2} \\ \times
\int 
\frac{ d^3 k_1}{ (2\pi)^3 2 k_1^0}
\frac{ d^3 k_2}{ (2\pi)^3 2 k_2^0}
(2\pi)^4 \delta^{(4)} (k_1 + k_2 - k)
\\ \times
\sum_{\text{spins}}
\left| \langle \gamma(k_1, \epsilon_1), \gamma(k_2, \epsilon_2) | T |  h''(k)  \rangle \right|^2,
\end{multline}
where $k^2 = m_{h''}^2$. Putting everything together we get for 
$d \sigma/ d m^2_{\gamma\gamma}$ with $h''$ in the phase-space region ${\cal B}$
\begin{multline} \label{eq3.8}
\frac{d \sigma}{ d m^2_{\gamma\gamma}} 
\left. \left( p(p_1) + p(p_2) \to (h'' \to \gamma\gamma) + X\right) \right|_{\cal B}
= \\
\frac{\Gamma(h'' \to \gamma\gamma)}{\Gamma_{h''}}
\frac{m_{h''} \Gamma_{h''} } {\pi} \\ \times
\frac{1}{(m^2_{\gamma\gamma} - m^2_{h''})^2 + m^2_{h''} \Gamma^2_{h''}}
\int_{\cal B} \frac{d^3 k}{k^0} f_{\text{inc}}(k). 
\end{multline}
Taking now ${\cal B}$ to be the total phase space for $h''$ production and integrating over $m^2_{\gamma\gamma}$ we find finally
\begin{multline} \label{eq3.9}
\sigma  \left( p(p_1) + p(p_2) \to (h'' \to \gamma\gamma) + X \right) =\\
\text{B}(h'' \to \gamma\gamma) \times \sigma \left( p(p_1) + p(p_2) \to h'' + X\right).
\end{multline}

The main approximations in~\eqref{eq3.9} are that we take the cross section for $h''$ production and the width
$\Gamma(h'' \to \gamma\gamma)$ as constant for an interval of size $\Gamma_{h''}$ around the nominal mass
value of $h''$, $m_{h''} = 95.4$~GeV. 
%Figure~\ref{fig10} shows that this is, in our case, a good approximation at the percent level.
In Appendix~\ref{appB} we present a discussion of the cross section on the l.h.s of~\eqref{eq3.9} without using these approximations.
Of course, once we have very precise data for $h'' \to \gamma\gamma$ we have to calculate also the true higher-order effects for the quantity~\eqref{eq3.9}.

In the MCPM' the main direct production mode of the $h''$ is the Drell-Yan reaction with a charm plus an
anticharm quark fusing to give $h''$; see Fig.~\ref{figDY}. 
\begin{figure}[!ht]
\centering
\begin{tikzpicture}
  \begin{feynman}
  %place first vertices
    \vertex (a) at (0,0);
     \vertex (i1) at (-1, 1) ;
     \vertex (i2) at (-1, -1) ;
 	 \vertex (f) at (1, 0){\(h''\)};
   \diagram* {
      (i1) -- [fermion, thick, edge label=\(c\)] (a),
      (a) -- [fermion, thick, edge label=\(\bar{c}\)] (i2),
      (a) -- [scalar, thick] (f),
 };
   \end{feynman}
\end{tikzpicture}
\caption{\label{figDY}
Partonic diagram for the 
Drell-Yan process for production of $h''$.}
\end{figure}
In addition there is gluon-gluon fusion giving $h''$ (Fig.~\ref{fig6}) and possibly also
$H^\pm$ production with the subsequent decays
\begin{equation} \label{3.10}
H^\pm \to h'' + W^\pm;
\end{equation}
see Fig.~\ref{fig7}. The couplings $H^\pm h'' W$ are given in Appendix~A of~\cite{Maniatis:2009vp}. 
Of course, the decays~\eqref{3.10} can only occur if $H^\pm$ are heavy enough, 
\begin{equation} \label{3.11}
m_{H^\pm} > m_{h''} + m_W = 175.8~\text{GeV}.
\end{equation}

\begin{figure}[!ht]
\centering
\begin{tikzpicture}
  \begin{feynman}
  %place first vertices
 %   \vertex (i1) at (0,1);
 %   \vertex (i2) at (0,-1);
     \vertex (a) at (2.5, 1) ;
     \vertex (b) at (2.5, -1) ;
     \vertex (c1) at (3.6, 0.5);
     \vertex (c2) at (3.6, -0.5);
     \vertex (c3) at (4.3, 0);
%     \vertex (X1) at (4, 1.5) ;
%     \vertex (X1a) at (4, 1.7) ;
%    \vertex (X1b) at (4, 1.3) ;
%     \vertex (X2) at (4, -1.5) ;
 %    \vertex (X2a) at (4, -1.3) ;
 %    \vertex (X2b) at (4, -1.7) ;
     \vertex (h) at (6,0) {\(h''\)};
    \diagram* {
%      (i1) -- [fermion, thick,edge label=\(p(p_1)\)] (a),
 %     (i2) -- [fermion, thick,edge label'=\(p(p_2)\)] (b),
      (a) -- [gluon, thick, edge label'=\(G\)] (c1),
      (b) -- [gluon, thick, edge label=\(G\)] (c2),
       (c1) -- [fermion, thick] (c3),
      (c3) -- [fermion, thick, edge label=\(q\)] (c2),
      (c2) -- [fermion, thick] (c1),
      (c3) -- [scalar, thick] (h),
%      (a) -- [plain, thick] (X1),
 %     (a) -- [plain, thick] (X1a),
  %    (a) -- [plain, thick] (X1b),
%      (b) -- [plain, thick] (X2),      
 %     (b) -- [plain, thick] (X2a),      
 %     (b) -- [plain, thick] (X2b),      
    };
 %  	\draw [fill, gray] (a) circle (.2);
% 	\draw [fill, gray] (b) circle (.2);
% 	\draw [fill] (c3) circle (.05);
  \end{feynman}
\end{tikzpicture}
\caption{\label{fig6} 
Partonic diagram for the production of $h''$ via gluon-gluon fusion with $q=c, s$.}
\end{figure}
\begin{figure}[!ht]
\centering
\begin{tikzpicture}
  \begin{feynman}
  %place first vertices
%    \vertex (i1) at (0,1);
%    \vertex (i2) at (0,-1);
     \vertex (a) at (2.0, 1) ;
     \vertex (b) at (2.0, -1) ;
     \vertex (c) at (3, 0);
     \vertex (d) at (4.5, 0);
     \vertex (f1) at (6, 1) {\(W^+\)};
     \vertex (f2) at (6, -1) {\(h''\)};
%     \vertex (X1) at (4, 1.5) ;
%     \vertex (X1a) at (4, 1.7) ;
%     \vertex (X1b) at (4, 1.3) ;
%     \vertex (X2) at (4, -1.5) ;
%     \vertex (X2a) at (4, -1.3) ;
%     \vertex (X2b) at (4, -1.7) ;
    \diagram* {
%      (i1) -- [fermion, thick,edge label=\(p(p_1)\)] (a),
%      (i2) -- [fermion, thick,edge label'=\(p(p_2)\)] (b),
      (a) -- [fermion, thick, edge label=\(c\)] (c),
      (b) -- [anti fermion, thick, edge label'=\(\bar{s}\)] (c),
      (c) -- [anti charged scalar, thick, edge label=\(H^+\), inner sep=5pt] (d),
      (d) -- [anti charged scalar, thick] (f1),
      (d) -- [scalar, thick] (f2),      
%      (a) -- [plain, thick] (X1),
%      (a) -- [plain, thick] (X1a),
%      (a) -- [plain, thick] (X1b),
%      (b) -- [plain, thick] (X2),      
%      (b) -- [plain, thick] (X2a),      
%      (b) -- [plain, thick] (X2b),      
    };
%   	\draw [fill, gray] (a) circle (.2);
% 	\draw [fill, gray] (b) circle (.2);
%	\draw [fill] (c) circle (.05);
  \end{feynman}
\end{tikzpicture}
\caption{\label{fig7}
Partonic diagram for Drell-Yan production of $H^+$ with its subsequent 
decay to $W^+ + h''$. The diagram for $H^-$ production and decay to $W^- + h''$ is analogous.}
\end{figure}

We have adapted the calculations of~\cite{Maniatis:2009vp}
to our case here, that is, $\sqrt{s} = 13$~TeV and $h'' = 95.4$~GeV. 
For the calculations of our Drell-Yan and gluon-gluon fusion processes,
Figs.~\ref{figDY}-\ref{fig7}, we use the quark and gluon distributions of the proton as given in 
\cite{Kovarik:2015cma} and implemented in \cite{Clark:2016jgm}, 
 the set ``nCTEQ15'' with 5 flavours at the one-loop order.
 Adopting the arguments presented in chapter~7.1 of~\cite{Ellis:1996mzs} we choose the 
 renormalization and factorization scales both equal to the mass of the~$h''$ boson.
We find the following cross sections
\begin{align} \label{eq3.12}
\sigma( p+p \to h'' + X) \bigg|_{\text{DY}} & = 17874.5~\text{pb},
\\
\label{eq3.13}
\sigma( p+p \to h'' + X) \bigg|_{\text{GG}} &= 210.2~\text{pb.}
\end{align}

We have made an estimate of the cross section for~$h''$ production through the~$H^\pm$ production with the subsequent decay~\eqref{3.10} and find that it gives at most of the order of 1\% of the DY cross section~\eqref{eq3.12}. 
Since this contribution is small and depends on the unknown masses of the~$H^\pm$ and $h'$ bosons, we neglect it in the following.

%The result for $H^\pm$ production with subsequent decay~\eqref{3.10} is shown in Fig.~\ref{fig8}. 
%Here also the value of the~$h'$ mass enters through the calculation of the total width~$\Gamma_{H^\pm}$ 
%and the branching fractions $\text{B}(H^\pm \to h'' + W^\pm)$. 
%We show the results for the, in our framework here, lowest possible value for $m_{h'}=m_{h''} = 95.4$~GeV 
%and for $m_{h'} = 1000$~GeV. 
%\begin{figure}[!ht]
%\centering
%\includegraphics[width=\columnwidth]{sigmaHbr}
%\caption{\label{fig8}
%The result for $\sum_{l=\pm} \sigma(p + p \to H^l +X) \times \text{B}(H^l \to h'' +W^l)$
%for $\sqrt{s} = 13$~TeV and $m_{h''} = 95.4$~GeV as function of the $H^\pm$ mass $m_{H^\pm}$. 
%In the lower curve the $h'$ mass is set to $m_{h'} = m_{h''} = 95.4$~GeV and in the upper curve
%to $m_{h'} = 1000$~GeV.
%}
%\end{figure}

\begin{figure}[!ht]
\centering
\includegraphics[width=\columnwidth]{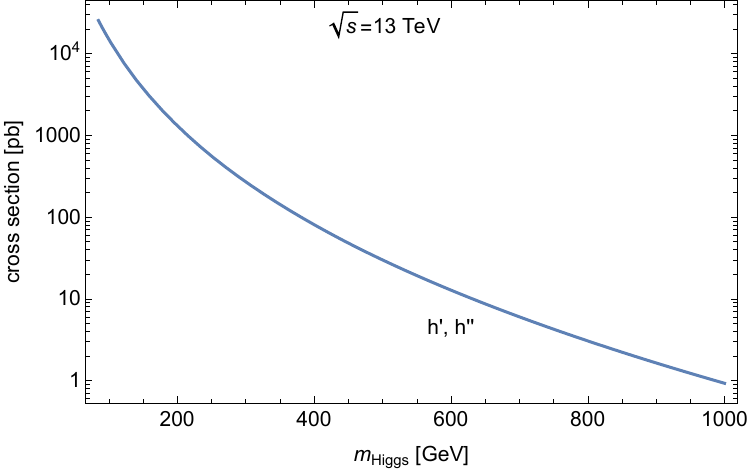}
\caption{\label{fig9}
Cross sections  for $h''$ and $h'$ production in $p p$ collisions at $\sqrt{s} = 13$~TeV via 
the Drell-Yan reactions $c\bar{c} \to h''$ and $c\bar{c} \to h'$.
}
\end{figure}

In Fig.~\ref{fig9} we show the cross sections for the Drell-Yan reactions for production of $h''$ and $h'$
as functions of the Higgs-boson masses for $\sqrt{s} = 13$~TeV. 

%%%%%%%%%%%%%%%%%%%%%%%%%%%%%%%%%%%%%%
%%%%%%%%%%%%%%%%%%%%%%%%%%%%%%%%%%%%%%
\section{Comparison with experiment and discussions}
\label{sec:comp}
\begin{figure}[!ht]
\centering
\includegraphics[width=\columnwidth]{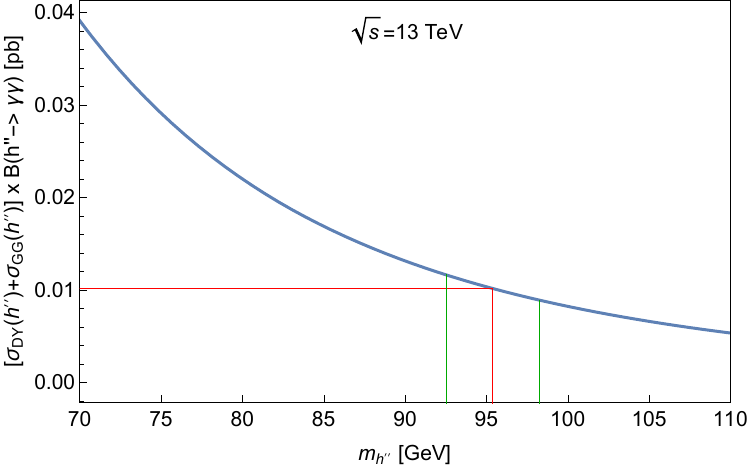}
\caption{\label{fig10}
The product $\big[\sigma_{\text{DY}}(p + p \to h'' + X) +\sigma_{\text{GG}}(p + p \to h'' + X) \big]$
$\times 
\text{B}(h'' \to \gamma + \gamma)$ for $\sqrt{s} = 13$~TeV as function of the $h''$ mass is shown.
The red lines correspond to $m_{h''}=95.4$~GeV where the CMS experiment sees an enhancement; see Fig.~5 of~\cite{CMS:2023yay}. We also indicate $m_{h''}\pm \Gamma_{h''}/2$ by the green lines. }
\end{figure}

First we discuss $\sigma(p + p \to h'' + X) \times \text{B}(h'' \to \gamma+ \gamma)$.
From the CMS measurement we conclude that for 95.4~GeV there may be an
enhancement of this type of product of the order of 0.01-0.04~pb; see~\eqref{eq1.1} and Fig.~5 of~\cite{CMS:2023yay}.
For our $h''$ with $m_{h''} = 95.4$~GeV we find on the other hand, using only the Drell-Yan (DY) and the gluon-gluon (GG) fusion contributions to $\sigma(p + p \to h'' + X)$,
\begin{multline} \label{eq4.1}
\left[
	\sigma_{\text{DY}}(p + p \to h'' + X) +
	\sigma_{\text{GG}}(p + p \to h'' + X)
\right] \\
\times 
\text{B}(h'' \to \gamma + \gamma)
= 0.01~\text{pb};
\end{multline}
see \eqref{eq3.2}, \eqref{eq3.12}, and \eqref{eq3.13}. The product \eqref{eq4.1} is
shown as function of the $h''$ mass in Fig.~\ref{fig10}.
Taking the finite width of the boson~$h''$ into account we find that the r.h.s of~\eqref{eq4.1} 
may be reduced by up to~50\% depending on the acceptance interval in $m_{\gamma\gamma}$; see Appendix~\ref{appB}.
%
%
%is changed; see~\eqref{B1}, \eqref{B2} in Appendix~\ref{appB}. For this purpose, we sum the differential cross section $d \sigma(p(p_1)+p(p_2) \to \gamma \gamma + X)/d m_{\gamma\gamma}$ with respect to the invariant mass of the photon pair $m_{\gamma\gamma}$ in the narrow-width approximation. We find that the total cross section varies about a factor of two if we vary the summation interval. This means that the cross section $\sigma(p(p_1)+p(p_2) \to \gamma \gamma + X)$ remains at the same order of magnitude as the possible measurement at CMS, taking the width $\Gamma_{h''}$ into account.   

It is remarkable that choosing in the MCPM' the $h''$ mass $m_{h''} = 95.4$~GeV we find,
without any tuning of other parameters, a value~\eqref{eq4.1} which is 
quite well compatible with the possible
experimental result~\eqref{eq1.1}. 
Of course, this experimental {\em result} could be 
a statistical fluctuation. Then the above {\em agreement} with our theory would be completely
fortuitous. But let us, in the following, tentatively assume that, indeed, the above experimental result
indicates a new effect. We have then an experimental and some theoretical comments. \\

In~\cite{CMS:2023yay} the analysis was done for a Standard-Model-like Higgs boson. We do not
know if and how the results of~\cite{CMS:2023yay} would change if this assumption is dropped and if the
production modes for our boson $h''$ would be considered. 
It would be nice if the experimentalists could reanalyse their data in this way.

Our theoretical estimate for $\sigma_{\text{DY}}(p + p \to h'' + X)$ is almost certainly a lower limit for this cross section. 
We used only the leading-order Drell-Yan formula. For the usual Drell-Yan reactions,
production of the vector bosons $\gamma^*$, $Z$, $W^\pm$, it is known that higher order
corrections in $\alpha_s$, the strong coupling parameter, increase $\sigma_{\text{DY}}$; see for instance chapter~9.2 of~\cite{Ellis:1996mzs}. 
We expect, therefore, also for our case a similar situation. Furthermore,
for $H^\pm$ with a mass $m_{H^\pm} > 175.8$~GeV, $H^\pm$ production with subsequent decay 
$H^\pm \to h'' + W^\pm$~\eqref{3.10} will increase the $h''$ yield. 

\begin{figure}[!ht]
\centering
\includegraphics[width=\columnwidth]{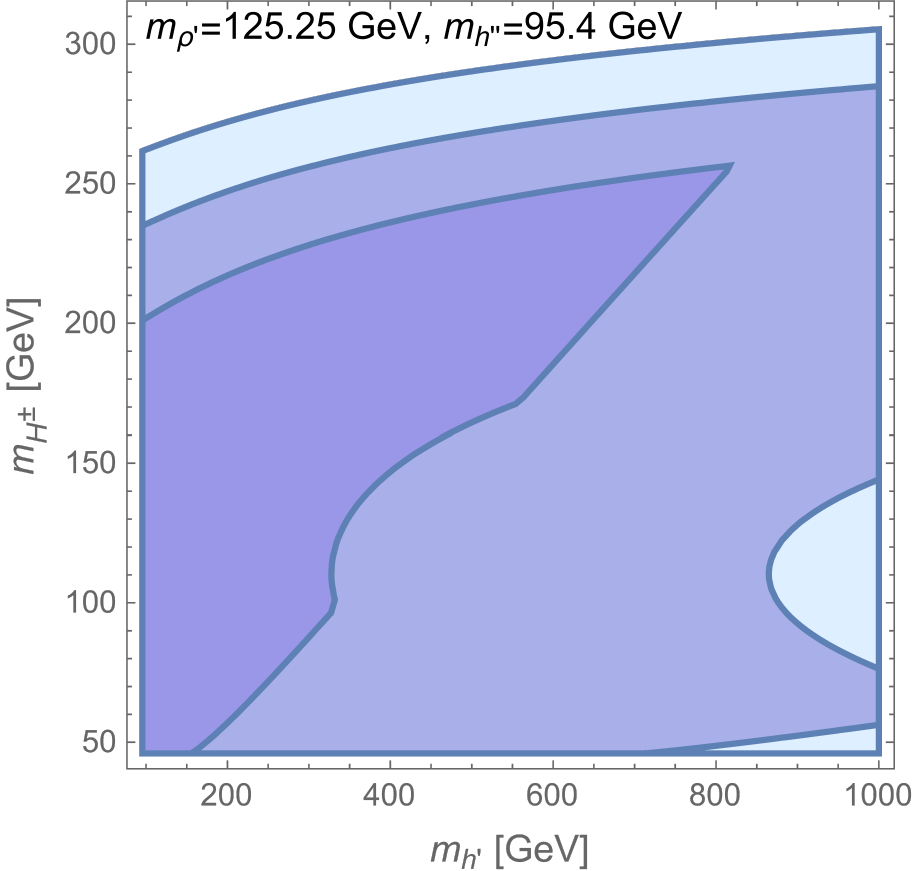}
\caption{
 \label{fig11} 
 The allowed regions in the $m_{h'}$--$m_{H^\pm}$ plane due
 to the values of the $S$, $T$, $U$ parameters~\eqref{eq4.2}.
 The masses of $\rho'$ and $h''$ are fixed to $m_{\rho'} = 125.25$~GeV and
 $m_{h''} = 95.4$~GeV. The shaded areas 
  correspond to the 1$\sigma$, 2$\sigma$, and 3$\sigma$ uncertainties of $S$, $T$, $U$ in \eqref{eq4.2}.
 }
\end{figure}

Encouraged by these considerations of a possible $h''$ boson at 95.4~GeV we are now looking at consequences for the Higgs bosons $h'$ and $H^\pm$ of the MCPM'. 
For this we use the method of~\cite{Maniatis:2011qu}. We consider the oblique parameters $S$, $T$, $U$
which have been computed and compared to the electroweak precision data~\cite{ParticleDataGroup:2024cfk} giving
\begin{equation} \label{eq4.2}
S= -0.02 \pm 0.10, \quad
T= 0.03 \pm 0.12, \quad
U= 0.01 \pm 0.11\;.
\end{equation}
The masses of the Higgs bosons of the MCPM' must respect the bounds on these parameters given in~\eqref{eq4.2}.
Now we fix in the MCPM' the masses of $\rho'$ and $h''$:
\begin{equation} \label{eq4.3}
m_{\rho'} = 125.25 \text{ GeV}, \quad
m_{h''} = 95.4 \text{ GeV}.
\end{equation}
We get then $1\sigma$, $2\sigma$, and $3\sigma$ regions in the $m_{h'}$--$m_{H^\pm}$ plane as shown in Fig.~\ref{fig11}.
Interestingly we find an allowed $3\sigma$ interval for the mass of the charged Higgs bosons of the order of
\begin{equation} \label{eq4.4}
45 \text{ GeV}  < m_{H^\pm} < 300 \text{ GeV}.
\end{equation}
The lower limit of $m_Z/2 \approx 45$~GeV comes from the fact that a decay $Z \to H^+H^-$ has not been observed.
The mass of $h'$ is allowed to be up to 1000~GeV.  But such high masses of $h'$ are unreasonable in the model. 
To justify our perturbative treatment we should probably require the parameters of the potential~\eqref{eq2.7}, \eqref{eq2.8}, to be not too large. But for $m_{h'}>1000$~GeV we find from~\eqref{eq2.11}
\begin{equation} \label{eq4.5}
\mu_1 - \mu_3 = \frac{m_{h'}^2}{2 v_0^2} > 8.2 \;.
\end{equation}
Thus, we consider $m_{h'} = 1000$~GeV as a generous upper limit  up to which the MCPM' still could make sense.
For very high values of the quartic parameters $\mu_1$, $\mu_2$, $\mu_3$ of the potential~\eqref{eq2.7}
we would have a strongly interacting Higgs sector where perturbative calculations would no longer be reliable. 

%%%%%%%%%%%%%%%%%%%%%%%%%%%%%%%%%%%%%%%%%%
%%%%%%%%%%%%%%%%%%%%%%%%%%%%%%%%%%%%%%%%%%
\section{Leptonic decays of $D_s^\pm$ and $D^\pm$ mesons in the MCPM'}
\label{sec:Ds}
In this section, we shall discuss aspects of flavor physics in the MCPM', that is, physics governed by the CKM matrix~V; see~\eqref{14.1}. 
In the strict symmetry limit we have from~\eqref{14.2} $V_\text{MCPM} = \unitmatrix_3$. 

In the MCPM' we generate deviations of the CKM matrix from $V=\unitmatrix_3$ as remnant effects from very high-mass-scale physics; see the discussion in Appendix~\ref{appA}, \eqref{A57}-\eqref{A70}. There we also give arguments that in our approach the CKM matrix~$V$ should have $|V_{33}|$ very close to~1, $|V_{13}|$, $|V_{23}|$, $|V_{31}|$, $|V_{32}|$ small, of order $\delta_\chi$ \eqref{eqA18}, \eqref{A19a}, and $|V_{12}|$, $|V_{21}|$ not particularly small. All this is indeed as observed experimentally; see chapter~12 of~\cite{ParticleDataGroup:2024cfk}.
 In contrast to this in the SM there is no prediction or estimate for the CKM matrix which could in principle be any unitary matrix. 

Keeping all this in mind we shall first study the leptonic decays of the charm-strange mesons~$D_s^\pm$, 
\begin{align}
&D_s^+(p) \to l^+(k_1, s_1) + \nu_l(k_2, s_2),\label{5.1}\\
&D_s^-(p) \to l^-(k_1, s_1) + \bar{\nu}_l(k_2, s_2), \label{5.2}
\quad l = e, \mu, \tau\;.
\end{align}
Here $p$, $k_1$, $k_2$ are the momenta and $s_1$, $s_2$ the helicities of the particles. In the SM these decays proceed through $W$-boson exchange. In the MCPM' we have in addition, for the muonic decay only, exchange of the $H^\pm$ bosons; see Fig.~\ref{fig200}.
%In the following we use for our calculations in the MCPM' the CKM matrix in the symmetry limit, $V=\unitmatrix_3$. As we shall see, this approximation works very well for the reactions~\eqref{5.1}, \eqref{5.2}. 

 \begin{figure}[!ht]
\centering
\begin{tabularx}{0.8\columnwidth}{p{0.1\columnwidth} p{0.9\columnwidth}}
(a)
&
\raisebox{-.45\height}{
\begin{tikzpicture}
  \begin{feynman}
  %place first vertices
    \vertex (v) at (0,0);
     \vertex (f1) at (1, 1){\(l^+(k_1, s_1)\)} ;
     \vertex (f2) at (1, -1){\(\nu_l(k_2, s_s)\)} ;
     \vertex (l2) at (-1.2,0);
     \vertex (l1) at (-2.2,0);
 	 \vertex (i) at (-3.7,0){\(D_s^+(p)\)};
   \diagram* {
      (f1) -- [fermion, thick] (v),
      (v) -- [fermion, thick] (f2),
      (v) -- [charged scalar, thick, edge label'=\(W^+\), inner sep = 5pt] (l2),
      (l1) -- [fermion, thick, half left, edge label=\(c\), inner sep = 5pt] (l2),
      (l2) -- [fermion, thick, half left, edge label=\(\bar{s}\), inner sep = 5pt] (l1),
      (l1) -- [charged scalar, thick] (i),
 };
   \end{feynman}
\end{tikzpicture}
}\\
(b)
&
\raisebox{-.45\height}{
\begin{tikzpicture}
  \begin{feynman}
  %place first vertices
    \vertex (v) at (0,0);
     \vertex (f1) at (1, 1){\(\mu^+(k_1, s_1)\)} ;
     \vertex (f2) at (1, -1){\(\nu_\mu(k_2, s_s)\)} ;
     \vertex (l2) at (-1.2,0);
     \vertex (l1) at (-2.2,0);
 	 \vertex (i) at (-3.7,0){\(D_s^+(p)\)};
   \diagram* {
      (f1) -- [fermion, thick] (v),
      (v) -- [fermion, thick] (f2),
      (v) -- [charged scalar, thick, edge label=\(H^+\), inner sep = 5pt] (l2),
      (l1) -- [fermion, thick, half left, edge label=\(c\), inner sep = 5pt] (l2),
      (l2) -- [fermion, thick, half left, edge label=\(\bar{s}\), inner sep = 5pt] (l1),
      (l1) -- [charged scalar, thick] (i),
 };
   \end{feynman}
\end{tikzpicture}
}
\end{tabularx}
\caption{\label{fig200} Diagrams of leading order for the leptonic decay of $D_s^\pm$.
In~(a) we have the $W$-exchange diagram for $l=e, \mu, \tau$, in~(b) the $H^\pm$ exchange diagram of the MCPM' which contributes only to the muonic decay. The diagrams for $D_s^-$ decay are analogous.
For the diagrams for~$D^+$ decay we replace the $s$ by the $d$ quark. 
}
\end{figure}

The decays~\eqref{5.1} and \eqref{5.2} have been studied thoroughly in the literature; see e.g.~\cite{Dobrescu:2008er,FlavourLatticeAveragingGroupFLAG:2021npn}
and the review~72 of~\cite{ParticleDataGroup:2024cfk}.
To leading order these low-energy processes are governed by effective Lagrangians. The exchange of the $W$ gives
us the following terms relevant for our discussions, 
\begin{multline} \label{5.3}
{\cal L}_\text{eff}^{(W)}(x) = 
- \frac{G_F}{\sqrt{2}} \sum_{l=e, \mu, \tau}
\bigg\{
\\
\bar{\nu}_l(x) \gamma^\lambda (1-\gamma_5) l(x)
\big[
 \bar{s}(x) V^*_{cs} \gamma_\lambda
(1-\gamma_5) c(x) 
\\
+ \bar{d}(x) V^*_{cd} \gamma_\lambda (1-\gamma_5) c(x)
+ \bar{s}(x) V^*_{us} \gamma_\lambda (1-\gamma_5) u(x)
\big]
\\
+
\bar{l}(x) \gamma^\lambda (1-\gamma_5) \nu_l(x) 
\big[
\bar{c}(x) V_{cs} \gamma_\lambda (1 - \gamma_5) s(x)
\\
+ \bar{c}(x) V_{cd} \gamma_\lambda (1-\gamma_5) d(x)
+ \bar{u}(x) V_{us} \gamma_\lambda (1-\gamma_5) s(x)
\big]
\bigg\}.
\end{multline}
Here $G_F$ is Fermi's constant and $V_{cs}$, $V_{cd}$, and $V_{us}$, are the appropriate elements of the Cabibbo-Kobayashi-Maskawa (CKM) matrix. The effective Lagrangian for $H^\pm$ exchange reads, using the couplings shown in~Fig.~\ref{fig100}
\begin{equation} \label{5.4}
{\cal L}_{\text{eff}}^{(H)}(x) = \frac{1}{m_{H^\pm}^2}  J^\dagger(x)J(x)\;.
\end{equation}
Here we have with $v_0$ from~\eqref{14a}
\begin{equation} \label{5.5a}
\begin{split}
J^\dagger(x) = &\frac{m_\tau}{\sqrt{2} v_0} \bar{\mu}(x) (1 - \gamma_5) \nu_\mu(x)\\
&- V_{cs}^* \big[ \frac{m_t - m_b}{\sqrt{2} v_0} \bar{s}(x) c(x)
+ \frac{m_t + m_b}{\sqrt{2} v_0} \bar{s}(x) \gamma_5 c(x) \big]\\
& - \frac{m_t}{\sqrt{2} v_0} V_{cd}^*  \big[ \bar{d}(x) c(x) + \bar{d}(x) \gamma_5 c(x) \big]
\\ &
+ \frac{m_b}{\sqrt{2} v_0} V_{us}^*  \big[ \bar{s}(x) u(x) - \bar{s}(x) \gamma_5 u(x) \big],\\
J(x) = &\frac{m_\tau}{\sqrt{2} v_0} \bar{\nu}_\mu(x) (1 + \gamma_5) \mu(x)\\
&- V_{cs} \big[ \frac{m_t - m_b}{\sqrt{2} v_0} \bar{c}(x) s(x)
- \frac{m_t + m_b}{\sqrt{2} v_0} \bar{c}(x) \gamma_5 s(x)\big]\\
& - \frac{m_t}{\sqrt{2} v_0} V_{cd}  \big[ \bar{c}(x) d(x) - \bar{c}(x) \gamma_5 d(x) \big]
\\ &
+ \frac{m_b}{\sqrt{2} v_0} V_{us}  \big[ \bar{u}(x) s(x) + \bar{u}(x) \gamma_5 s(x) \big].
\end{split}
\end{equation}
In~\eqref{5.5a} we have again only written down the terms of the currents~$J^\dagger$ and $J$ which are relevant for our discussion here. The complete currents are given in~\eqref{A77}; see the terms multiplying~$H^-$ and $H^+$, respectively. 

In the following we shall calculate the matrix elements for the decays~\eqref{5.1}
and \eqref{5.2} in leading order neglecting neutrino masses and mixings. The hadronic parts of these matrix elements are then QCD quantities which have to respect parity~($P$), time-reversal~($T$) and charge-conjugation~($C$) invariance. We have
\begin{multline} \label{5.6}
\langle 0 | \bar{c}(x) \gamma^\lambda \gamma_5 s(x) | D_s^-(p) \rangle
\\
= 
\langle 0 | \bar{s}(x) \gamma^\lambda \gamma_5 c(x) | D_s^+(p) \rangle
\\
= 
i p^\lambda f_{D_s} e^{-i p \cdot x},
\quad
f_{D_s}^* = f_{D_s}\;,
\end{multline}
and
\begin{multline} \label{5.7}
\langle 0 | \bar{c}(x)  \gamma_5 s(x) | D_s^-(p) \rangle
\\
= 
\langle 0 | \bar{s}(x) \gamma_5 c(x) | D_s^+(p) \rangle
\\
= 
-i m_{D_s} \tilde{f}_{D_s} e^{-i p \cdot x},
\quad
\tilde{f}_{D_s}^* = \tilde{f}_{D_s}\;.
\end{multline}
For the following it is convenient to define
\begin{equation} \label{5.7a}
r_{D_s} = \left(\frac{f_{D_s}}{\tilde{f}_{D_s}}\right)^{1/2}\;.
\end{equation}

The decay constant $f_{D_s}$ has been computed by lattice QCD methods.
A recent review~\cite{FlavourLatticeAveragingGroupFLAG:2021npn} gives
\begin{equation}
f_{D_s} = (249.9 \pm 0.5 )~\text{MeV};
\end{equation}
see also Table~72.4 of~\cite{ParticleDataGroup:2024cfk}.
To estimate $\tilde{f}_{D_s}$ we can use the divergence relation (see \cite{Dobrescu:2008er})
\begin{equation} \label{5.9}
i \frac{\partial}{\partial x^\lambda} \bar{c}(x) \gamma^\lambda \gamma_5 s(x)
=
- (m_c+m_s) \bar{c}(x) \gamma_5 s(x)\;.
\end{equation}
This gives from~\eqref{5.6} and \eqref{5.7}
\begin{equation} \label{5.10}
\tilde{f}_{D_s} = \frac{m_{D_s}}{m_c + m_s} f_{D_s}\;.
\end{equation}
The problem with~\eqref{5.10} is that we do not know at which scale
we should take the quark masses~$m_c$ and $m_s$. Using the \msbar masses (see Table~\ref{tablemass})
as quoted in PDG \cite{ParticleDataGroup:2024cfk} we get for $f_{D_s}/\tilde{f}_{D_s}$
and $r_{D_s}= (f_{D_s}/\tilde{f}_{D_s})^{1/2}$
\begin{equation} \label{5.11}
\frac{f_{D_s}}{\tilde{f}_{D_s}} = \frac{m_c + m_s}{m_{D_s}} = 0.69, \quad
r_{D_s} =  \sqrt{ \frac{f_{D_s}}{\tilde{f}_{D_s}}} = 0.83\;.
\end{equation}

But maybe we should, for the low energy decays~\eqref{5.1} and \eqref{5.2}, rather
use constituent quark masses~$m_c$ and $m_s$. Therefore, we give now another
estimate for~$\tilde{f}_{D_s}$. We split the Dirac-field operators into upper and
lower components, 
\begin{equation} \label{5.12}
c(x) = \begin{pmatrix} \varphi_c(x)\\ \chi_c(x) \end{pmatrix}, \quad
s(x) = \begin{pmatrix} \varphi_s(x)\\ \chi_s(x) \end{pmatrix}\;.
\end{equation}
We have then
\begin{equation} \label{5.13}
\begin{split}
& \bar{c}(x) \gamma^0 \gamma_5 s(x) = \varphi_c^\dagger(x) \chi_s(x)
+ \chi_c^\dagger(x) \varphi_s(x), \\
& \bar{c}(x) \gamma_5 s(x) = \varphi_c^\dagger(x) \chi_s(x)
- \chi_c^\dagger(x) \varphi_s(x)\;.
\end{split}
\end{equation}
For non-relativistic antiquark $\bar{c}$ and quark $s$ in the $D^-_s$ state at rest,
$p = p_R = (m_{D_s}, 0, 0, 0)^\trans$, we expect that only the terms $\chi_c^\dagger \varphi_s$ in~\eqref{5.13} will contribute to 
$\langle 0 | \bar{c}(x)  \gamma^0 \gamma_5 s(x) | D_s^-(p_R) \rangle$
and
$\langle 0 | \bar{c}(x) \gamma_5 s(x) | D_s^-(p_R) \rangle$.
Thus, we expect to have 
\begin{multline} \label{5.14}
\langle 0 | \bar{c}(x)  \gamma^0 \gamma_5 s(x) | D_s^-(p_R) \rangle\\
\approx
\langle 0 | \chi_c^\dagger(x) \varphi_s(x) | D_s^-(p_R) \rangle\\
\approx
- \langle 0 | \bar{c}(x) \gamma_5 s(x) | D_s^-(p_R) \rangle\;.
\end{multline}
From~\eqref{5.6} and \eqref{5.7} we get then
\begin{equation} \label{5.15}
\tilde{f}_{D_s} \approx f_{D_s}, \quad
\frac{f_{D_s}}{\tilde{f}_{D_s}} \approx 1, \quad
r_{D_s} = \sqrt{\frac{f_{D_s}}{\tilde{f}_{D_s}}} \approx 1\;.
\end{equation}

In the following we shall, for our estimates, assume the ratio~$r_{D_s}$~\eqref{5.7a} to be in the range
\begin{equation} \label{5.16}
0.8 \le r_{D_s} \le 1.2\;.
\end{equation}

Now we can calculate the $T$-matrix element for the decay~\eqref{5.1} where we take the final state as
\begin{equation} \label{5.17}
| l^+(k_1, s_1), \nu_l(k_2, s_2)\rangle = b^\dagger_l(k_1, s_1) a^\dagger_{\nu_l}(k_2, s_2) |0\rangle\;.
\end{equation}
Here $b^\dagger$ and $a^\dagger$ are the appropriate creation operators. 
By~\eqref{5.17} we have fixed the phase of the final state. In the SM we have only the $W$-exchange diagram, Fig.~\ref{fig200}~(a), and we get the well known result
\begin{multline} \label{5.18}
\langle l^+(k_1, s_1), \nu_l(k_2, s_2) | T | D_s^+(p) \rangle_{\text{SM}}
\\
= i \frac{G_F}{\sqrt{2}} V^*_{cs} m_l f_{D_s} \bar{u}_{\nu_l}(k_2, s_2) 
(1 + \gamma_5) v_l(k_1, s_1), \\
l = e, \mu, \tau\;.
\end{multline}
For the decay rates this gives
\begin{multline} \label{5.19}
\Gamma(D_s^+ \to l^+ \nu_l)_{\text{SM}} \\
= 
\frac{G_F^2}{8\pi} |f_{D_s}|^2 |V_{cs}|^2 m_{D_s} m_l^2 
\left( 1- \frac{m_l^2}{m_{D_s}^2} \right)^2,\quad
 l = e, \mu, \tau\;.
\end{multline}
The factors $m_l$ in \eqref{5.18} and $m_l^2$ in \eqref{5.19} are the famous 
helicity-suppression factors which lead to the hierarchy
\begin{multline} \label{5.20}
\Gamma(D_s^+ \to e^+ \nu_e)_{\text{SM}} \\ \ll 
\Gamma(D_s^+ \to \mu^+ \nu_\mu)_{\text{SM}} \\ \ll
\Gamma(D_s^+ \to \tau^+ \nu_\tau)_{\text{SM}}\;.
\end{multline}
We note that the SM decay rates~\eqref{5.19} contain the factor $|V_{cs}|^2$ which,
a priori, is not fixed in the SM, except for $0 \le |V_{cs}|^2 \le 1$. Thus, these decay rates are one possibility to {\em determine} $|V_{cs}|$, and this is indeed what is done. Taking our leading order calculation~\eqref{5.19} for the $\tau^+ \nu_\tau$ decay we get, with the masses and the total width~$\Gamma_{D_s}$ from PDG~\cite{ParticleDataGroup:2024cfk},
\begin{multline} \label{5.21}
\text{B}(D_s^+ \to \tau^+ \nu_\tau)_{\text{SM}} 
\\
= |V_{cs}|^2 \frac{G_F^2}{8\pi}
|f_{D_s}|^2 m_{D_s} m_\tau^2 \left(1- \frac{m_\tau^2}{m_{D_s}^2}\right)^2 
\Gamma^{-1}_{D_s}
\\
= |V_{cs}|^2 \cdot 5.51 \times 10^{-2} \;.
\end{multline}
Comparing this with the experimental value from PDG~\cite{ParticleDataGroup:2024cfk},
\begin{equation} \label{5.21a}
\text{B}(D_s^+ \to \tau^+ \nu_\tau)_{\text{exp}} = (5.36 \pm 0.10) \times 10^{-2}\;,
\end{equation}
we get, taking the central value from~\eqref{5.21a},
\begin{equation} \label{5.22}
|V_{cs}|= 0.990\;.
\end{equation}
This is close to the value $|V_{cs}| = 0.975 \pm 0.006$ 
which is obtained taking into account experimental results for 
leptonic plus semileptonic~$D_s$ decays and theoretical corrections in
the SM calculations; see~(12.10) of the review~12 of~\cite{ParticleDataGroup:2024cfk}.

Finally we consider the ratio
\begin{equation} \label{5.23}
R(D_s)_{\tau \mu} = \frac{\Gamma(D_s^+ \to \tau^+ \nu_\tau)}{\Gamma(D_s^+ \to \mu^+ \nu_\mu)}\;.
\end{equation}
In the SM this ratio is fixed, as there~$|V_{cs}|^2$ drops out:
\begin{equation} \label{5.24}
\left. R(D_s)_{\tau \mu} \right|_{\text{SM}} =
\frac{ m_\tau^2 \left(1- \frac{m_\tau^2}{m_{D_s}^2} \right)^2}
{ m_\mu^2 \left(1- \frac{m_\mu^2}{m_{D_s}^2} \right)^2}\;.
\end{equation}
With the masses of the particles from PDG~\cite{ParticleDataGroup:2024cfk} we
get
\begin{equation} \label{5.25}
\left. R(D_s)_{\tau \mu} \right|_{\text{SM}} = 9.75\;;
\end{equation}
see (72.27) of~\cite{ParticleDataGroup:2024cfk}. The experimental values 
obtained from the new PDG results~\cite{ParticleDataGroup:2024cfk} are
\begin{equation} \label{69a}
\begin{split}
&B(D_s^+ \to e^+ \nu_e) < 8.3 \cdot 10^{-5},\\
&B(D_s^+ \to \mu^+ \nu_\mu) = (5.35 \pm 0.12) \cdot 10^{-3},\\
&B(D_s^+ \to \tau^+ \nu_\tau) = (5.36 \pm 0.10) \cdot 10^{-2}.
\end{split}
\end{equation}
From this we find
\begin{equation} \label{5.25a}
\left. R(D_s)_{\tau \mu} \right|_{\text{exp}} = 10.02 \pm 0.29\;.
\end{equation}

Now we come to the decays~\eqref{5.1} and \eqref{5.2} in the MCPM',
where
%,as mentioned above, we set for the CKM matrix $V=\unitmatrix_3$. 
%In our framework this means neglecting some remnants of physics at a high mass scale in~$V$. 
%In the MCPM' 
we have for the decay
$D_s^+ \to l^+ \nu_l$ with $l=e, \tau$ the prediction 
as in~\eqref{5.18}, \eqref{5.19}.
% but setting $V_{cs}=1$. Thus, we predict in the MCPM'
%\begin{equation} \label{5.26}
%\Gamma(D_s^+ \to \tau^+ \nu_\tau)_{\text{MCPM'}} 
%=  \frac{G_F^2}{8\pi}
%|f_{D_s}|^2 m_{D_s} m_\tau^2 \left(1- \frac{m_\tau^2}{m_{D_s}^2}\right)^2\;,
%\end{equation}
%\begin{equation} \label{5.27}
%\text{B}(D_s^+ \to \tau^+ \nu_\tau)_{\text{MCPM'}} 
%=5.51 \times 10^{-2}\;.
%\end{equation}
%This agrees with experiment~\eqref{5.21a} to within~3.6\% or~1.7~$\sigma$. Note that our MCPM' relations~\eqref{5.26}, \eqref{5.27} indeed hold in reality up to small corrections, at the percent level. In our framework such corrections would be due to remnants of high-mass-scale physics.  

However, for
the decay $D_s^+ \to \mu^+ \nu_\mu$ we have in the MCPM' both,
the $W$- and the $H$-exchange diagram; see Fig.~\ref{fig200}. This gives
\begin{multline} \label{5.28}
\langle \mu^+(k_1, s_1), \nu_\mu(k_2, s_2) | T | D_s^+(p) \rangle_{\text{MCPM'}}
\\
= i \frac{G_F}{\sqrt{2}}  V_{cs}^* m_\mu f_{D_s} 
\left[ 1 - \frac{\bar{m}^2_{D_s}}{r_{D_s}^2 m_{H^\pm}^2} \right]
\\ 
\times
\bar{u}_{\nu_\mu}(k_2, s_2) 
(1 + \gamma_5) v_\mu(k_1, s_1)\;, 
\end{multline}
where~$r_{D_s}$ is defined in~\eqref{5.7a} and 
\begin{equation} \label{5.29}
\bar{m}^2_{D_s} = \frac{ m_\tau (m_t + m_b)}{2 v_0^2}
\frac{ \sqrt{2} m_{D_s}}{G_F m_\mu} = (76.5~\text{GeV})^2\;.
\end{equation}
From~\eqref{5.28} we get for the ratio~$R(D_s)_{\tau\mu}$~\eqref{5.23}
in the MCPM'
\begin{equation} \label{5.30}
\left. R(D_s)_{\tau\mu}\right|_{\text{MCPM'}} =
\frac{ m_\tau^2 \left(1-\frac{m_\tau^2}{m_{D_s}^2}\right)^2}
{ m_\mu^2 \left(1-\frac{m_\mu^2}{m_{D_s}^2}\right)^2}
\left( 1 - \frac{\bar{m}^2_{D_s}}{r_{D_s}^2 m_{H^\pm}^2} \right)^{-2}\;.
\end{equation}

%Note that in the SM $\Gamma(D_s^+ \to \tau^+ \nu_\mu)$ contains the free
%parameter $|V_{cs}|$ and $R_{\tau \mu}$ is fixed; see~\eqref{5.19} and \eqref{5.24},
%respectively. In the MCPM' the situation is reversed:
%$\Gamma(D_s^+ \to \tau^+ \nu_\mu)$ is fixed, and $R_{\tau \mu}$ contains the free parameter $m_{H^\pm}$; see~\eqref{5.26} and \eqref{5.30}, respectively. 

\begin{figure}[!ht]
\centering
\includegraphics[width=\columnwidth]{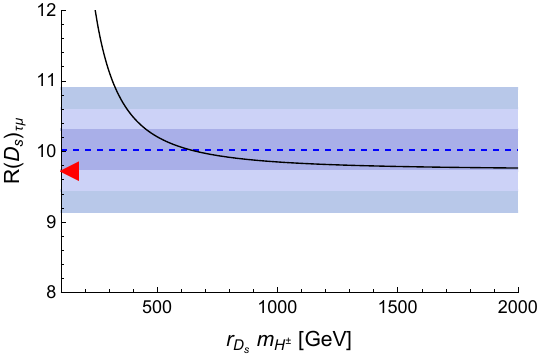}
\caption{\label{plotRDs} 
The ratio $R(D_s)_{\tau \mu}$~\eqref{5.23}. Shown are the experimental value~\eqref{5.25a} with its one, two, and three sigma deviations as shaded regions, and the prediction of the MCPM'~\eqref{5.30} as function of $r_{D_s} m_{H^\pm}$~(full line). A small triangle at the ordinate shows the prediction of the SM~\eqref{5.25}.
}
\end{figure}

In Fig.~\ref{plotRDs} we show $\left. R(D_s)_{\tau \mu} \right|_{\text{MCPM'}}$
as function of $r_{D_s} m_{H^\pm}$ and the experimental value 
$\left. R(D_s)_{\tau \mu} \right|_{\text{exp}}$ from \eqref{5.25a}. We find 
agreement of the MCPM' prediction with the experimental value at the $1\sigma$, 
$2\sigma$, and $3\sigma$ level for
\begin{equation} \label{5.31}
\begin{split}
&r_{D_s} \cdot m_{H^\pm} \ge 454 \text{ GeV \quad at 1} \sigma\text{ level,}\\
&r_{D_s} \cdot m_{H^\pm} \ge 374 \text{ GeV \quad at 2} \sigma\text{ level,}\\
&r_{D_s} \cdot m_{H^\pm} \ge 327 \text{ GeV \quad at 3} \sigma\text{ level.}\\
\end{split}
\end{equation}
With the ratio $r_{D_s}$ in the range~\eqref{5.16} this translates to
\begin{equation} \label{5.32}
\begin{split}
&m_{H^\pm} \ge 378 \text{ GeV \quad at 1} \sigma\text{ level,}\\
&m_{H^\pm} \ge 312 \text{ GeV \quad at 2} \sigma\text{ level,}\\
&m_{H^\pm} \ge 272 \text{ GeV \quad at 3} \sigma\text{ level.}\\
\end{split}
\end{equation}

Next we discuss the leptonic decays of the $D^\pm$ mesons
\begin{equation} \label{77.1}
\begin{split} 
&D^+(p) \to l^+(k_1, s_1) + \nu_l(k_2, s_2),\\
&D^-(p) \to l^-(k_1, s_1) + \bar{\nu}_l(k_2, s_2),
\end{split}
\end{equation}
where $l = e, \mu, \tau$. Our analysis here is completely analogous to the one for the~$D_s$ leptonic decays.
We define the decay constants through
\begin{align}
\label{77.3}
& \langle 0 | \bar{d}(x) \gamma^\lambda \gamma_5 c(x) | D^+(p) \rangle =
i p_\lambda f_D e^{-ipx},\\
\label{77.4}
&\langle 0 | \bar{d}(x) \gamma_5 c(x) | D^+(p) \rangle =
-i m_D \tilde{f}_D e^{-ipx},
\end{align}
and
\begin{equation} \label{77.5}
r_D = \left( \frac{f_D}{\tilde{f}_D} \right)^{1/2}. 
\end{equation}
In the following we shall assume, as in the~$D_s$ case, 
\begin{equation} \label{77.5a}
0.8 \le r_D \le 1.2\;.
\end{equation}
Furthermore we define the analog of $\bar{m}_{D_s}^{2}$ \eqref{5.29} as
\begin{equation} \label{77.6}
\bar{m}_D^2 = \frac{m_\tau m_t}{2 v_0^2} \frac{\sqrt{2} m_D}{G_F m_{\mu}} = (73.8~\text{GeV})^2.
\end{equation}
We get then as in~\eqref{5.18} and \eqref{5.28}
\begin{multline} \label{77.7}
\langle l^+(k_1, s_1), \nu_l(k_2, s_2) | T | D^+(p) \rangle_{\text{MCPM'}}
\\
= i \frac{G_F}{\sqrt{2}}  V_{cd}^* m_l f_{D} 
\left[ 1 - \delta_{l\mu} \frac{\bar{m}^2_D}{r_D^2 m_{H^\pm}^2} \right]
\\
\times
\bar{u}_{\nu_l}(k_2, s_2) 
(1 + \gamma_5) v_l(k_1, s_1)\;,\\
l = e, \mu, \tau; \quad \delta_{l\mu} = \begin{cases} 1, &\text{for }  l = \mu,\\ 0, &\text{for } l = e, \tau. \end{cases}
\end{multline}
This gives for the decay rates
\begin{multline} \label{77.8}
\Gamma(D^+ \to l^+ \nu_l)_{\text{MCPM'}} \\
= 
\frac{G_F^2}{8\pi} |f_{D}|^2 |V_{cd}|^2 m_{D} m_l^2 
\left( 1- \frac{m_l^2}{m_D^2} \right)^2
\\ \times
\left( 1- \delta_{l\mu} \frac{\bar{m}_D^2} {r_D^2 m^2_{H^\pm}} \right)^2.
\end{multline}
We define the ratio analogous to~\eqref{5.23} as
\begin{equation} \label{77.9}
R(D)_{\tau \mu} = \frac{\Gamma(D^+ \to \tau^+ \nu_\tau)}{\Gamma(D^+ \to \mu^+ \nu_\mu)}
=
\frac{B(D^+ \to \tau^+ \nu_\tau)}{B(D^+ \to \mu^+ \nu_\mu)}.
\end{equation}
From~\eqref{77.8} we get
\begin{equation} \label{86a}
\left.R(D)_{\tau \mu}\right|_{\text{MCPM'}} 
= \frac{m_\tau^2 (m_D^2 - m_\tau^2 )^2} {m_\mu^2 (m_D^2 - m_\mu^2 )^2}
\left( 1 - \frac{\bar{m}_D^2}{r_D^2 m_{H^\pm}^2} \right)^{-2}
\end{equation}
with~$r_D$ from~\eqref{77.5}, \eqref{77.5a}, and $\bar{m}_D^2$ from~\eqref{77.6}.

\begin{figure}[!ht]
\centering
\includegraphics[width=\columnwidth]{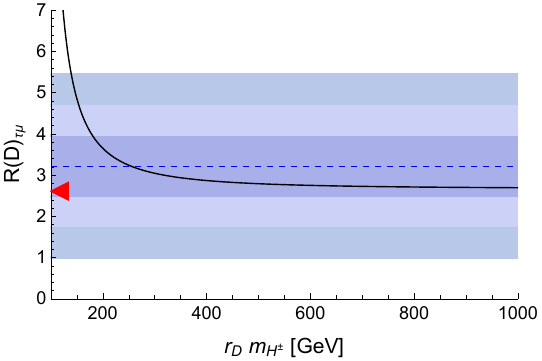}
\caption{\label{plotRD} 
The ratio $R(D)_{\tau \mu}$~\eqref{77.9}. Shown are the experimental value~\eqref{77.11} with its one, two, and three sigma deviations as shaded regions, and the prediction of the MCPM'~\eqref{77.8} as function of $r_{D} m_{H^\pm}$~(full line). A small triangle at the ordinate shows the prediction of the SM~\eqref{77.12}.
}
\end{figure}

The experimental numbers for the leptonic decay branching ratios of the~$D^+$ mesons are~\cite{ParticleDataGroup:2024cfk}
\begin{equation} \label{77.10}
\begin{split}
&B(D^+ \to e^+\nu_e) < 8.8 \cdot 10^{-6},\\
&B(D^+ \to \mu^+\nu_\mu) = (3.74 \pm 0.17) \cdot 10^{-4},\\
&B(D^+ \to \tau^+\nu_\tau) = (1.20 \pm 0.27) \cdot 10^{-3}.
\end{split}
\end{equation}
This gives
\begin{equation} \label{77.11}
\left. R(D)_{\tau\mu}\right|_{\text{exp}} = 3.21 \pm 0.74.
\end{equation}
For the SM we get from~\eqref{86a}, setting~$\bar{m}_D^{2} = 0$,
\begin{equation} \label{77.12}
\left. R(D)_{\tau\mu}\right|_{\text{SM}} = 2.66.
\end{equation}
The prediction of our MCPM' for $R(D)_{\tau\mu}$ is shown as function of $r_D m_{H^\pm}$ in~Fig.~\ref{plotRD}. 
 Comparing with the experimental number~\eqref{77.11} we get the following limits for 
$r_D m_{H^\pm}$:
\begin{equation} \label{77.13}
\begin{split}
&r_{D} \cdot m_{H^\pm} \ge 174 \text{ GeV \quad at 1} \sigma\text{ level,}\\
&r_{D} \cdot m_{H^\pm} \ge 148 \text{ GeV \quad at 2} \sigma\text{ level,}\\
&r_{D} \cdot m_{H^\pm} \ge 134 \text{ GeV \quad at 3} \sigma\text{ level.}
\end{split}
\end{equation}
Using~\eqref{77.5a} this translates to 
\begin{equation} \label{77.14}
\begin{split}
&m_{H^\pm} \ge 145 \text{ GeV \quad at 1} \sigma\text{ level,}\\
&m_{H^\pm} \ge 123 \text{ GeV \quad at 2} \sigma\text{ level,}\\
&m_{H^\pm} \ge 112 \text{ GeV \quad at 3} \sigma\text{ level.}\\
\end{split}
\end{equation}

Finally we have a short look at the~$K^\pm$ leptonic decays
\begin{equation} 
\label{77.15}
\begin{split}
&K^+(p) \to l^+(k_1, s_1) + \nu_l(k_2, s_2),\\
&K^-(p) \to l^-(k_1, s_1) + \bar{\nu}_l(k_2, s_2),\\
& \quad l = e, \mu.
\end{split}
\end{equation}
Here we have for the decay constants
\begin{equation} \label{77.16}
\begin{split}
&\langle 0 | \bar{s}(x) \gamma_\lambda \gamma_5 u(x) | K^+(p) \rangle =
i p_\lambda f_K e^{-ipx},\\
&\langle 0 | \bar{s}(x) \gamma_5 u(x) | K^+(p) \rangle =
-i m_K \tilde{f}_K e^{-ipx},\\
&\quad \text{with } f_K = f_K^*, \quad \tilde{f}_K = \tilde{f}_K^*\;.
\end{split}
\end{equation}
We set again
\begin{equation} \label{77.17}
r_K = \left( \frac{f_K}{\tilde{f}_K} \right)^{1/2}
\end{equation}
and define 
\begin{equation} \label{77.18}
\bar{m}_K^2 = \frac{m_\tau m_b}{2 v_0^2} \frac{\sqrt{2} m_K}{G_F m_{\mu}} = (5.90~\text{GeV})^2,
\end{equation}
and a ratio analogous to~\eqref{5.23} and \eqref{77.9}
\begin{equation} \label{77.19}
R(K)_{e \mu} = \frac{\Gamma(K^+ \to e^+ \nu_e)}{\Gamma(K^+ \to \mu^+ \nu_\mu)}
=
\frac{B(K^+ \to e^+ \nu_e)}{B(K^+ \to \mu^+ \nu_\mu)}.
\end{equation}
The SM and the MCPM' values for $R(K)_{e\mu}$, neglecting SM radiative corrections, are
\begin{align} \label{77.20}
&\left. R(K)_{e\mu} \right|_{\text{SM}} =
\frac{m_e^2(m_K^2 - m_e^2)^2}{m_\mu^2(m_K^2 - m_\mu^2)^2},
\\
 \label{77.21}
&\left. R(K)_{e\mu} \right|_{\text{MCPM'}} =
\left. R(K)_{e\mu} \right|_{\text{SM}}
\left( 1 - \frac{\bar{m}^2_K}{r_K^2 m^2_{H^\pm}} \right)^{-2} .
\end{align}
The experimental values from~\cite{ParticleDataGroup:2024cfk} for the branching fractions are 
\begin{equation} \label{77.22}
\begin{split}
&B(K^+ \to e^+ \nu_e) = (1.582 \pm 0.007) \cdot 10^{-5}, \\
&B(K^+ \to \mu^+ \nu_\mu) = (63.56 \pm 0.11) \cdot 10^{-2}.
\end{split}
\end{equation}
This leads to
\begin{equation} \label{77.23}
\left. R(K)_{e\mu}\right|_{\text{exp}} = (2.489 \pm 0.012) \cdot 10^{-5},
\end{equation}
which is a result with an error of 4.8\textperthousand,. 
On the other hand, our MCPM' predicts from~\eqref{77.21}, 
taking $r_K m_{H^\pm} = 300$~GeV, an increase of $R(K)_{e\mu}$ from the SM value of the
order of 0.8\textperthousand. 
We see that the present experimental precision for $R(K)_{e\mu}$ does not allow us to draw any conclusions for $m_{H^\pm}$ from this quantity. \\

%%%%%%%%%%%%%%%%%%%%%%%%%%%%%%%%%%%%%%%%%%
%%%%%%%%%%%%%%%%%%%%%%%%%%%%%%%%%%%%%%%%%%
\section{Conclusions}
\label{sec:con}
%\todo{ Investigations into CP symmetries within the two-Higgs-Doublet Model (2HDM) have identified a distinctive CP symmetry which is not equivalent to the standard CP symmetry. This unique symmetry naturally results in the Maximally CP-symmetric Model (MCPM), offering an explanation for family replication~\cite{Maniatis:2007de}.}
In this paper we have extended the maximally-CP-symmetric model, MCPM, from~\cite{Maniatis:2007de} to a realistic model which we call MCPM'. In the MCPM only the fermions of the third family have masses, the fermions of  the second and first family remain massless. Clearly, this is only a first approximation to what is observed in nature. In the present work we discuss in detail the generation of masses for the first and second family and of a nontrivial CKM matrix in the MCPM'. The key idea is to assume the existence of high-mass (of order~10~TeV) fields~$\chi_i(x)$, ($i=1,2$) and $\zeta(x)$. 
The effective interactions generated by these fields for the ``low energy'' LHC regime give us the fermion masses for the second and first generation.
 We also get an interesting method to generate EWSB; see~\eqref{eqA11}, \eqref{eqA12}. 
 The CKM matrix which we obtain has automatically some salient features as observed in experiment (see \eqref{A69} and \eqref{A70}). 
All this is explained in our Appendix~\ref{appA}. 
The Higgs sector of the MCPM' is as in the MCPM; see~\cite{Maniatis:2007de}--\cite{Maniatis:2009by}.

In our study, we have investigated the implications of the hypothesis that the 
possible
 anomaly observed by CMS~\cite{CMS:2023yay} in the $\gamma\gamma$ yield at 95.4~GeV may be attributed to the production and decay of the $h''$ boson of the MCPM’.
We found 
from our calculation for the
product $\sigma(p + p \to h'' + X) \times \text{B}(h'' \to \gamma + \gamma) \approx 0.01$~pb, to be 
compared with the experimental result of 0.01--0.04~pb for this quantity. 
We argued that our above result should be considered rather as a lower limit from theory. 
How could 
experimentalists explore the consequences of this finding further?\\

The MCPM' gives also predictions for the decay $h'' \to \mu^+ \mu^-$ from the coupling shown in 
Fig.~\ref{fig3},
\begin{equation} \label{eq5.1}
\begin{split}
&\Gamma(h'' \to \mu^+ \mu^-) = 0.198~\text{MeV},\\
&\text{B}(h'' \to \mu^+ \mu^-) = 3.44 \times 10^{-5}\;.
\end{split}
\end{equation}
Therefore, we have the prediction for $m_{h''} = 95.4$~GeV
\begin{multline} \label{eq5.2}
\left[\sigma_{\text{DY}}(p + p \to h'' + X) +\sigma_{\text{GG}}(p + p \to h'' + X) \right] \\
\times 
\text{B}(h'' \to \mu^+ + \mu^- )=0.62~\text{pb}.
\end{multline}
However, $m_{h''} = 95.4$~GeV is rather close to the $Z$ mass $m_Z \approx 91.2$~GeV and the decay $Z \to \mu^+ \mu^-$ will present a large source of background. 
Maybe one could consider the ratio of $\mu^+\mu^-$ and $e^+ e^-$ production.
For $\mu^+\mu^-$ the boson $h''$ will contribute but {\em not} for $e^+e^-$. 

Finally, the best way to establish the $h''$ would be to observe its main decay $h'' \to c\bar{c}$, that is, $h'' \to$ charm jet plus anticharm jet. 
Clearly, this represents a formidable challenge 
from the experimental side
given the 
background from QCD $c\bar{c}$ jets and $Z \to c\bar{c}$ decays. 
The possibility to observe the production of $h''$ with subsequent decay to charm jet plus anticharm jet will depend on how well charm tagging can be done and what resolution of the invariant mass of two such tagged jets can be achieved. 
Such an experimental study is far beyond the scope of this article, 
which primarily focuses on the theoretical predictions.
In this context it may be 
very advantageous to consider diffractive production of the $h''$ in $pp$ collisions,
where the $h''$ is produced in pomeron-pomeron ($\mathbbm{P}$-$\mathbbm{P}$) collisions.
This type of reactions was first discussed in~\cite{Schafer:1990fz}, where two processes were identified,
the inclusive and exclusive one; see Figs.~1 and 2 of~\cite{Schafer:1990fz}. In the exclusive case we have
$\mathbbm{P} + \mathbbm{P} \to h''$, that is,
\begin{equation} \label{eq5.3}
p + p \to p + h'' + p\;.
\end{equation}
In the inclusive case we have $\mathbbm{P} + \mathbbm{P} \to h''+X$, where $X$ would consist of two (presumably rather low energy)
gluonic jets close to beam direction
\begin{equation} \label{eq5.4}
p + p \to p + h'' + X + p\;.
\end{equation}
%In~\cite{Schafer:1990fz} numerical estimates for Higgs production via the inclusive reaction~\eqref{eq5.4} were presented.
%Numerical results for the
%exclusive Higgs-production reaction~\eqref{eq5.3} 
Of course, in~\cite{Schafer:1990fz} diffractive production of the SM Higgs particle~$H$ was considered. For this we have to replace~$h''$ by $H$ in~\eqref{eq5.3} and \eqref{eq5.4}.
In~~\cite{Schafer:1990fz} numerical estimates for~$H$ production via the inclusive reaction analogous to~\eqref{eq5.4} were presented. Numerical results for exclusive~$H$ production (see~\eqref{eq5.3} with $h''$ replaced by~$H$)
were first presented in \cite{Bialas:1991wj}.
 For reviews of exclusive SM-Higgs production we 
refer to~\cite{Albrow:2000na,Bartels:2006ea,Harland-Lang:2013zoo,LHCHiggsCrossSectionWorkingGroup:2013rie}.

Studying the reactions~\eqref{eq5.3} and \eqref{eq5.4} will require measurement of the outgoing protons, that is, it
would need
forward detectors. 
Then one could make a missing-mass analysis for detecting the possible production of $h''$ and with a central 
detector the remnant $X$ in~\eqref{eq5.4} and
 the decays of $h''$ could be analysed. Presumably, in such a setup there will be much less background than
in normal inclusive $h''$ production, where it will be accompanied by a large number of SM particles. 
The exclusive process~\eqref{eq5.3} also offers the possibility to check the pseudoscalar couplings to fermions of the boson~$h''$. 
For this 
one could use the methods explained in~\cite{Kaidalov:2003fw,Lebiedowicz:2013ika}.\\

In~Sec.~\ref{sec:Ds} we have discussed MCPM' effects in the leptonic decays of the charm-strange mesons~$D_s^\pm$
and the charm mesons~$D^\pm$. In particular, we studied the ratios~$R(D_s)_{\tau\mu}$ from~\eqref{5.23}
and $R(D)_{\tau\mu}$ from~\eqref{77.9}
of the decay rates of~$D^\pm_s$ and $D^\pm$ to $\tau \nu_\tau$ and
$\mu \nu_\mu$ leptons; see Fig.~\ref{plotRDs} and Fig.~\ref{plotRD}. 
From this we concluded that the charged Higgs-bosons~$H^\pm$ of the MCPM' should have a mass  $m_{H^\pm} \gtrsim 270$~GeV; see~\eqref{5.32}.
On the other hand, our study of the oblique parameters $S$, $T$, $U$ in Fig.~\ref{fig11}
suggested $m_{H^\pm} \lesssim 300$~GeV.
Thus, we have in the MCPM' the prediction that there should be a pair of charged Higgs bosons~$H^\pm$ at a mass around $m_{H^\pm} \approx 300$~GeV.
The main production mode of $H^\pm$ at the LHC should be the Drell-Yan reaction with $c\bar{s}$ and $s\bar{c}$ fusion, see Fig.~\ref{fig7}, where also the subsequent decay $H^+ \to W^+ h''$ is indicated. 

Production and decay properties of the~$H^\pm$ bosons have been calculated in~\cite{Maniatis:2009vp} in the framework of the MCPM. In the MCPM', our framework here, the gross features of the $H^\pm$ properties are as in the MCPM but in detail there are important differences. This can be seen from~\eqref{A77} where the couplings of~$H^\pm$ to fermions in the MCPM' are given. The detailed analysis of the~$H^\pm$ production and decay properties in the MCPM' goes beyond the scope of our present paper and will be the topic of a forthcoming study. 

Finally we can ask about effects of the MCPM' in other areas of particle physics, for instance,
involving top and bottom quarks and $\tau$ leptons. As stated in~\cite{Maniatis:2007de} and 
recalled here in Sec.~\ref{sec:basics} these fermions have in the MCPM' exactly the same couplings to the MCPM' boson $\rho'$ as they have in the SM to the SM Higgs boson. 
Also, $\rho'$ behaves very much like the SM Higgs boson; see the list of Feynman rules in Appendix~A of~\cite{Maniatis:2009vp}.
Thus, MCPM' effects involving $t$, $b$, and $\tau$ fermions distinct from SM effects are expected to be very small. Next we can ask about MCPM' effects for particles containing charm quarks. Exchange of the $h''$ will contribute to the $c\bar{c}$ potential; see Fig.~\ref{fig12}.
\begin{figure}[!ht]
\centering
\begin{tikzpicture}
  \begin{feynman}
  %place first vertices
    \vertex (a) at (0,0);
     \vertex (i1) at (-1, 1) ;
     \vertex (i2) at (-1, -1){\(c\)} ;
     \vertex (b) at (2,0) ;
 	 \vertex (f1) at (3, -1){\(\bar{c}\)};
 	 \vertex (f2) at (3, 1) ;
   \diagram* {
%      (i1) -- [anti fermion, thick, edge label=\(f\)] (a),
		(i2) -- [fermion, thick] (a),
      	(a) -- [fermion, thick] (i1),
      (a) -- [scalar, thick, edge label=\(h''\)] (b),
		(f2) -- [fermion, thick] (b),
      	(b) -- [fermion, thick] (f1)
 };
   \end{feynman}
\end{tikzpicture}
\caption{\label{fig12}
Diagram for $h''$ exchange contributing to the $c\bar{c}$ potential.
}
\end{figure}
However, due to~\eqref{2.16}
this will be a weak correction to the strong $c\bar{c}$ potential, similar to $Z$ exchange. 
Thus, it will be hard to observe this $h''$ effect.
 In decays like 
\begin{equation} \label{5.5}
J/\Psi  \to \text{ hadrons} + (h'' \to \mu^+\mu^-) 
\end{equation}
the virtual $h''$ has a coupling of the order of the electromagnetic one to the charm quarks;
see~\eqref{2.16}, \eqref{2.17}. 
But its coupling to $\mu^+\mu^-$ is (see Fig.~\ref{fig3})
\begin{equation}
c''_\mu = - \frac{m_\tau}{v_0} = - \frac{1.777~\text{GeV}}{246~\text{GeV}} = -7.2 \cdot 10^{-3}.
\end{equation}
Thus, $|c''_\mu |$ is much smaller than $e$~\eqref{2.17}, and decays like~\eqref{5.5} are expected to be extremely rare, much rarer than weak decays of the $J/\Psi$!

In~\cite{LEPWorkingGroupforHiggsbosonsearches:2003ing}
 a summary of the SM Higgs search at LEP was presented. A possible signal of a SM like Higgs boson at a mass about 98~GeV was reported in Higgs strahlung in events of the type $Z(H \to b\bar{b})$ and $Z(H \to \tau^+ \tau^-)$. The statistical significance of the effects was given to be around two standard deviations; see Fig.~7 
of~\cite{LEPWorkingGroupforHiggsbosonsearches:2003ing}. In our model it would be difficult to accommodate such effects.

We emphasize that in our paper we explore the predictions of the model MCPM'. We did not try to fit all very small, possibly beyond SM, effects reported by experiments. It is not unheard of that some small effects also disappear with time.
 But let us also emphasize that in the MCPM' we still can add some small couplings of the Higgs field~$\varphi_2$ to fermions, but such couplings would be suppressed by our small parameter~$\delta_\chi$~\eqref{eqA18}, \eqref{A19a}; see~\eqref{A71}--\eqref{A73} and the paragraph following it. 
 Indeed, if our MCPM' is correct, it will be a nice task for experimentalists to pin down the couplings of the Higgs-boson~$\varphi_2$ to the fermions. 
 The MCPM' makes the prediction that the main couplings are those of~\eqref{A77}. 

To conclude: motivated by the findings of CMS~\cite{CMS:2023yay} we have explored the consequences of having 
the pseudoscalar Higgs boson $h''$ of the MCPM' at a mass of 95.4~GeV. 
We have presented arguments from a study of the oblique parameters and of~$D_s$ and $D$ leptonic decays that the charged Higgs-boson pair~$H^\pm$ of the MCPM' should have a mass~$m_{H^\pm}$ around~300~GeV.
We have then given a number of
predictions which experimentalists working at LHC should be able to check. 
Thus, as it should be, 
our theory is falsifiable. 
% =============================================================================
 \acknowledgments
 We thank Christian Schwanenberger for very useful discussions and information concerning the CMS experimental results.
 We also thank A.~Szczurek for discussions and comments.
 This work is supported by Chile ANID FONDECYT project 1200641. 
 
 %%%%%%%%%%%%%%%%%%%%%%%%%%%%%%%%%%%%%%%%%%
%%%%%%%%%%%%%%%%%%%%%%%%%%%%%%%%%%%%%%%%%%
\appendix
\section{Generation of masses for the fermions of the first and second generation
and of the CKM matrix in the MCPM'}
\label{appA}

We start from the MCPM where the Higgs potential is given in~\eqref{eq2.7}
which we now write as
 \begin{equation} \label{eqA1}
 V_{\text{MCPM}} = \xi_0^{\text{orig}} K_0 + \eta_{00} K_0^2 + \tvec{K}^\trans E \tvec{K}\;,
 \end{equation}
with $\xi_0^{\text{orig}}$ the original parameter. At this starting point only the fermions of the third generation, $t$, $b$, $\tau$, have non-zero masses.  

To illustrate our ideas and methods we shall first discuss how we generate nonzero masses for the fermions of the second family. In a second step we shall then show how we generate masses for the second and first families together with the nontrivial CKM matrix. 
We introduce high-mass fields
$\chi_i(x)$, $(i=1,2)$ and $\zeta(x)$. Under the electroweak gauge group \eweakgroup~the fields $\chi_i$ are assumed to be doublets with hypercharge $y=1/2$, the real field $\zeta$ is assumed to be a singlet with $y=0$. We introduce the following interaction Lagrangian which has only dimension four terms
\begin{equation} \label{A2}
\tilde{\cal L}(x) = \sum_{i=1}^2 \bigg[ \tilde{J}_i^\dagger(x) \chi_i(x) + \chi_i^\dagger(x) \tilde{J}_i(x) \bigg],
\end{equation}
where
\begin{equation} \label{eqA3}
\begin{split}
\tilde{J}_1(x) =& a \zeta^2(x) \varphi_1(x) - b^{(\mu)} \bar{\mu}_R(x) 
\begin{pmatrix} \nu_{\mu L}(x)\\ \mu_L(x) \end{pmatrix}
\\ &
+ b^{(c)} \epsilon 
\begin{pmatrix} \bar{c}_L(x) \\ \bar{s}_L(x) \end{pmatrix} c_R(x)
- b^{(s)} \bar{s}_R(x)
\begin{pmatrix} c_L(x)\\ s_L(x) \end{pmatrix},\\
\tilde{J}_2(x) =& a \zeta^2(x) \varphi_2(x)\;.
\end{split}
\end{equation}
Here $a$, $b^{(\mu)}$, $b^{(c)}$, and $b^{(s)}$, are real, positive, constants,
\begin{equation} \label{eqA4}
\mu_L(x) = \frac{1}{2} (1-\gamma_5) \mu(x),\quad
\mu_R(x) = \frac{1}{2} (1+\gamma_5) \mu(x),
\end{equation}
and similarly for the other fermion fields, and 
\begin{equation} \label{eqA5}
\epsilon = \begin{pmatrix} \phantom{+}0 & 1\\ -1& 0 \end{pmatrix}.
\end{equation}

Now we assume that the $\chi_i(x)$ fields have a common mass $m_\chi$ which is very large
and that their propagator has the following form at low energies
\begin{equation} \label{eqA5a}
\langle 0 | T\big(\chi_i(x_1) \chi_j^\dagger(x_2)\big) | 0 \rangle \to
-\frac{i}{m_\chi^2} \delta^{(4)}(x_1 - x_2) \delta_{ij} \unitmatrix_2\;.
\end{equation}
For the field $\zeta(x)$ we assume that it has a vacuum-expectation value
\begin{equation} \label{eqA6}
\langle 0 | \zeta(x) | 0 \rangle = \zeta_0\;,
\end{equation}
and that also the field $\zeta'(x) = \zeta(x) - \zeta_0$ has a very high mass.
Then we can integrate out the $\chi_i$ fields and replace $\zeta(x)$ by $\zeta_0$. 
The resulting effective Lagrangian at LHC energies reads as follows
\begin{equation} \label{eqA7}
{\cal L}_{\text{eff}}(x) = \sum_{l=1}^3 {\cal L}_{\text{eff}}^{(l)}(x)\;,
\end{equation}
where
\begin{equation} \label{eqA8}
\begin{split}
{\cal L}_{\text{eff}}^{(1)}(x) =  &\frac{(a \zeta_0^2)^2}{m_\chi^2}
\big[ \varphi_1^\dagger(x) \varphi_1(x) + \varphi_2^\dagger(x) \varphi_2(x) \big]\\
=  &\frac{(a \zeta_0^2)^2}{m_\chi^2} K_0(x)\;,
\end{split}
\end{equation}
\begin{multline} \label{eqA9}
{\cal L}_{\text{eff}}^{(2)}(x) =  -\frac{a \zeta_0^2}{m_\chi^2} \varphi_1^\dagger(x)
\big[ b^{(\mu)} \bar{\mu}_R(x) \begin{pmatrix} \nu_{\mu L}(x)\\ \mu_L(x) \end{pmatrix}\\
- b^{(c)} \epsilon \begin{pmatrix} \bar{c}_L(x)\\ \bar{s}_L(x) \end{pmatrix} c_R(x)
+b^{(s)} \bar{s}_R(x) \begin{pmatrix} c_L(x)\\ s_L(x) \end{pmatrix} \big] + h.c.\;,
\end{multline}
\begin{multline} \label{eqA10}
{\cal L}_{\text{eff}}^{(3)}(x) = \frac{1}{m_\chi^2} 
\big[ b^{(\mu)} \begin{pmatrix} \bar{\nu}_{\mu L}, & \bar{\mu}_L(x) \end{pmatrix} \mu_R(x)
\\
- b^{(c)} \bar{c}_R(x) \begin{pmatrix} c_L(x), & s_L(x) \end{pmatrix} \epsilon^\trans
+ b^{(s)} \begin{pmatrix} \bar{c}_L(x), & \bar{s}_L(x) \end{pmatrix} s_R(x) \big]
\\ \times
\big[ b^{(\mu)} \bar{\mu}_R(x) \begin{pmatrix} \nu_{\mu L}(x)\\ \mu_L(x) \end{pmatrix}
- b^{(c)} \epsilon \begin{pmatrix} \bar{c}_L(x) \\ \bar{s}_L(x) \end{pmatrix} c_R(x)
\\
+ b^{(s)} \bar{s}_R(x) \begin{pmatrix} c_L(x)\\ s_L(x) \end{pmatrix} \big]\;.
\end{multline}

Let us discuss these three terms ${\cal L}_{\text{eff}}^{(l)}$, $l=1,2,3$ in turn.
The first term ${\cal L}_{\text{eff}}^{(1)}$ gives a contribution to the Higgs potential~\eqref{eqA1}
changing $\xi_0^{\text{orig}}$ to $\xi_0$ to be used in~\eqref{eq2.7}:
\begin{equation} \label{eqA11}
\xi_0^{\text{orig}} \to \xi_0 = \xi_0^{\text{orig}} - \frac{(a \zeta_0^2)^2}{m_\chi^2}\;.
\end{equation}
EWSB requires $\xi_0<0$; see~\eqref{eq2.10}. An attractive possibility is, therefore,
to have $ \xi_0^{\text{orig}}=0$ and 
\begin{equation} \label{eqA12}
\xi_0 =  - \frac{(a \zeta_0^2)^2}{m_\chi^2}\;.
\end{equation}
Then EWSB and the connected spontaneous breaking of the $\text{CP}_g^{(i)}$ 
symmetry would originate from effects of the high mass physics.
We have then from~\eqref{eq2.11} and \eqref{eq4.3}
\begin{equation} \label{eqA13}
m_{\rho'}^2 = 2(- \xi_0) = 2 \frac{a \zeta_0^2}{m_\chi^2} a \zeta_0^2 = (125.25~\text{GeV})^2\;.
\end{equation}

The term ${\cal L}_{\text{eff}}^{(2)}$ gives mass terms for the fermions $\mu$, $c$, and $s$. 
Indeed, after EWSB we get 
\begin{equation} \label{eqA14}
\varphi_1(x) = \frac{v_0}{\sqrt{2}} \begin{pmatrix} 0 \\ 1 + \frac{\rho'(x)}{v_0} \end{pmatrix},
\end{equation}
\begin{multline} \label{eqA15}
{\cal L}_{\text{eff}}^{(2)}(x) = -\frac{a \zeta_0^2}{m_\chi^2}  \frac{v_0}{\sqrt{2}}
\big( 1 + \frac{\rho'(x)}{v_0} \big)
\big[ b^{(\mu)} \bar{\mu}(x) \mu(x)
\\
 + b^{(c)} \bar{c}(x) c(x) + b^{(s)} \bar{s}(x) s(x) \big]\;.
\end{multline}
From this we can read off the masses
\begin{multline} \label{eqA16}
m_\mu = \frac{a \zeta_0^2}{m_\chi^2} \frac{v_0}{\sqrt{2}} b^{(\mu)}, \quad
m_c = \frac{a \zeta_0^2}{m_\chi^2} \frac{v_0}{\sqrt{2}} b^{(c)}, \\
m_s = \frac{a \zeta_0^2}{m_\chi^2} \frac{v_0}{\sqrt{2}} b^{(s)}\;.
\end{multline}
Our Higgs field $\rho'(x)$ couples to these fermions with constants proportional to the masses. This is exactly as for the SM Higgs particle and this is indeed what is observed in experiments~\cite{ATLAS:2022vkf}. 

The term ${\cal L}_{\text{eff}}^{(3)}$, \eqref{eqA10}, gives effective four-fermion interactions but suppressed by small couplings proportional to $m_\chi^{-2}$. 

Let us consider a numerical example, in order to see if our ansatz could make sense. 
We choose 
\begin{equation} \label{eqA17}
b^{(c)} = 1
\end{equation}
and get from~\eqref{eqA16}
\begin{equation} \label{eqA18}
\frac{a \zeta_0^2}{m_\chi^2} = \frac{m_c \sqrt{2}}{v_0} = 7.3 \times 10^{-3}\;.
\end{equation}
In the following this parameter will play an important role. Therefore, we give it an extra name
\begin{equation} \label{A19a}
\delta_\chi = \frac{a \zeta_0^2}{m_\chi^2}\;.
\end{equation}
For numerical results we shall use the value $\delta_\chi = 7.3 \cdot 10^{-3}$ from~\eqref{eqA18} as an example.

Then~\eqref{eqA13} gives
\begin{equation} \label{eqA19}
a \zeta_0^2 = \frac{m_{\rho'}^2}{2 \delta_\chi} = 1.07~\text{TeV}^2\;.
\end{equation}
From \eqref{eqA16}, \eqref{eqA18}, and \eqref{eqA19}, we find
\begin{equation} \label{eqA20}
m_\chi = 12.1~\text{TeV}, \quad
\frac{\sqrt{2}}{v_0 \delta_\chi} = 0.788~\text{GeV}^{-1}\;,
\end{equation}

\begin{equation} \label{eqA21}
\begin{split}
&b^{(s)} = m_s
\frac{\sqrt{2}}{v_0 \delta_\chi}  = 7.3 \times 10^{-2}\;,\\
&b^{(\mu)} = m_\mu
\frac{\sqrt{2}}{v_0 \delta_\chi}  = 8.3 \times 10^{-2}\;.
\end{split}
\end{equation}
Here we use the mass values $m_c$, $m_s$, $m_\mu$ as given by PDG~\cite{ParticleDataGroup:2024cfk}; see Table~\ref{tablemass}. All the parameters of our model calculated in this example in~\eqref{eqA17}-\eqref{eqA21} look quite reasonable to us.

Now we extend our considerations in order to show how we can generate masses for the second and first families and a nontrivial CKM matrix. It will be convenient to give leptons and quarks a family index,
\begin{alignat}{4} 
&l_1 = e, \quad &&l_2 = \mu, \quad &&l_3= \tau, \nonumber\\
&u_1 = u, \quad &&u_2 = c, \quad &&u_3 = t, \label{A23}\\
&d_1 = d, \quad &&d_2 = s, \quad &&d_3 = b \nonumber\;.
\end{alignat}
We shall denote the $\mathit{SU}(2)_L$ doublets by
\begin{equation} \label{A24}
\begin{pmatrix} u_{\alpha L}\\ d'_{\alpha L} \end{pmatrix}, \qquad
\begin{pmatrix} u'_{\alpha L}\\ d_{\alpha L} \end{pmatrix}, \qquad \alpha = 1,2,3,
\end{equation}
where $u_{\alpha}$ and $d_{\alpha}$ will be the mass eigenfields.

Our starting point is the Yukawa Lagrangian of the MCPM as given in (120) and (121) of~\cite{Maniatis:2007de}.
\begin{align}\label{A25}
{\cal L}_{\mathrm{Yuk, MCPM}} = \notag\\ 
  -c^{(1)}_{l\,3} & \;\Bigg\{
    \bar{l}_{3\,R}\,\varphi_1^\dagger
    \begin{pmatrix} \nu_{3\,L} \\ l_{3\,L} \end{pmatrix}
    -\bar{l}_{2\,R}\,\varphi_2^\dagger
    \begin{pmatrix} \nu_{2\,L} \\ l_{2\,L} \end{pmatrix}
    \Bigg\}
\notag\\
  +c^{(1)}_{u\,3} &\;\Bigg\{
    \bar{u}''_{3\,R}\,\varphi_1^\trans\,\epsilon
    \begin{pmatrix} u'_{3\,L} \\ d_{3\,L} \end{pmatrix}
    -\bar{u}_{2\,R}\,\varphi_2^\trans\,\epsilon
    \begin{pmatrix} u_{2\,L} \\ d'_{2\,L} \end{pmatrix}
    \Bigg\}
\notag\\
  -c^{(1)}_{d\,3} &\;\Bigg\{
    \bar{d}_{3\,R}\,\varphi_1^\dagger
    \begin{pmatrix} u'_{3\,L} \\ d_{3\,L} \end{pmatrix}
    -\bar{d}_{2\,R}\,\varphi_2^\dagger
    \begin{pmatrix} u'_{2\,L} \\ d_{2\,L} \end{pmatrix}
    \Bigg\}
		+ h.c.
\end{align}
Here we have
\begin{equation} \label{A26}
\begin{split}
&c^{(1)}_{l\,3} = \frac{\sqrt{2}}{v_0} m_\tau^{\text{orig}} \approx \frac{\sqrt{2}}{v_0} m_\tau = 1.02 \cdot 10^{-2},\\
&c^{(1)}_{u\,3} = \frac{\sqrt{2}}{v_0} m_t^{\text{orig}} \approx \frac{\sqrt{2}}{v_0} m_t = 0.99,\\
&c^{(1)}_{d\,3} = \frac{\sqrt{2}}{v_0} m_b^{\text{orig}} \approx \frac{\sqrt{2}}{v_0} m_b = 2.40 \cdot 10^{-2}.\\
\end{split}
\end{equation}
At this level we have 
$u'_{\alpha L} = u_{\alpha L}$,
$u''_{\alpha R} = u_{\alpha R}$,
$d'_{\alpha L} = d_{\alpha L}$,
$d'_{\alpha R} = d_{\alpha R}$,
but for the following it is important to keep the freedom to have $u'_{\alpha L}$ and $u''_{\alpha R}$ different from
$u_{\alpha L}$ and $u_{\alpha R}$, ($\alpha = 1,2,3$), respectively. 
As we shall see, the original values $m_\tau^{\text{orig}}$, $m_t^{\text{orig}}$, and $m_b^{\text{orig}}$ will turn out to be slightly different from the corresponding final, experimental, values.

Now we make an ansatz for the coupling of our high-mass fields $\chi_i$ and $\zeta$ as in \eqref{A2} and \eqref{eqA3} but with different currents
$\tilde{J}_i \to J_i$, ($i = 1,2$). We set again 
\begin{equation} \label{A27}
\tilde{\cal L}(x) = \sum_{i=1}^2 \bigg[ J_i^\dagger(x) \chi_i(x) + \chi_i^\dagger(x) J_i(x) \bigg],
\end{equation}
where now
\begin{align} 
&J_1(x) = a \zeta^2(x) \varphi_1(x) + J_1^{(f)}(x),\nonumber\\
& \qquad J_1^{(f)}(x)= - \sum_{\alpha=1}^3 b_{l \alpha}^{(1)} \bar{l}_{\alpha R}(x) 
\begin{pmatrix} \nu_{\alpha L}(x)\\ l_{\alpha L}(x) \end{pmatrix}
\nonumber\\ 
& \qquad
+ \sum_{\alpha, \beta=1}^3 \left(\tilde{b}_{u\; \alpha\beta}^{(1)}\right)^* \epsilon
\begin{pmatrix} \bar{u}'_{\beta L}(x)\\ d_{ \beta L}(x) \end{pmatrix} u''_{\alpha R}(x)
\label{A28}\\ 
& \qquad
-
\sum_{\alpha=1}^3 b_{d \alpha}^{(1)} \bar{d}_{\alpha R}(x) 
\begin{pmatrix} u'_{ \alpha L}(x)\\ d_{\alpha L}(x) \end{pmatrix}, \nonumber\\ 
\label{A28a}
&J_2(x) = a \zeta^2(x) \varphi_2(x) + J_2^{(f)}(x).
\end{align}
Here the superscript $(f)$ stands for the fermionic part of the currents and $a$,
$b_{l\, \alpha}^{(1)}$, $b_{d\, \alpha}^{(1)}$, $\tilde{b}_{u\, \alpha\beta}^{(1)}$ ($\alpha, \beta = 1,2,3$) are constants with 
\begin{equation} \label{A29}
a > 0, \quad
b_{l\, \alpha}^{(1)} > 0, \quad
b_{d\, \alpha}^{(1)} > 0, \quad
\tilde{b}_{u\, \alpha\beta}^{(1)} \text{ complex}\;.
\end{equation}
The form of the current~$J_2^{(f)}(x)$ will be specified later. From~\eqref{A27}--\eqref{A28a} we get the effective action at low energies in complete analogy to~\eqref{eqA6}--\eqref{eqA10},
\begin{equation} \label{A31}
{\cal L}_{\text{eff}}(x) = 
{\cal L}_{\text{eff}}^{(1)}(x) + {\cal L}_{\text{eff}}^{(2)}(x) + {\cal L}_{\text{eff}}^{(3)}(x) \;,
\end{equation}
where
\begin{equation} \label{A32}
{\cal L}_{\text{eff}}^{(1)}(x) = \frac{(a \zeta_0^2)^2}{m_\chi^2} K_0(x)\;,
\end{equation}
\begin{equation} \label{A33}
\begin{split} 
&{\cal L}_{\text{eff}}^{(2)}(x) = {\cal L}_{\text{eff}}^{(2,1)}(x) + {\cal L}_{\text{eff}}^{(2,2)}(x) \;,\\
&{\cal L}_{\text{eff}}^{(2,1)}(x) = \delta_\chi \left\{ \varphi_1^\dagger(x) J_1^{(f)}(x) + h.c. \right\}\;,\\
&{\cal L}_{\text{eff}}^{(2,2)}(x) = \delta_\chi \left\{ \varphi_2^\dagger(x) J_2^{(f)}(x) + h.c.\right\}\;,
\end{split}
\end{equation}
\begin{equation} \label{A34}
{\cal L}_{\text{eff}}^{(3)}(x) =  \frac{1}{m_\chi^2} 
\sum_{i=1}^2 {J_i^{(f)}}^\dagger (x) J_i^{(f)} (x)\;.
\end{equation}

The term~${\cal L}_{\text{eff}}^{(1)}$ is as in~\eqref{eqA8} and with~\eqref{eqA11} and \eqref{eqA12} 
will produce EWSB, as discussed there. The term~${\cal L}_{\text{eff}}^{(3)}$ produces four-fermion interactions but suppressed by the large mass-squared scale~$m_\chi^2$. 
The term~${\cal L}_{\text{eff}}^{(2)}$, together with ${\cal L}_{\text{Yuk, MCPM}}$ from~\eqref{A25} gives the Yukawa interaction
\begin{equation} \label{A35}
{\cal L}_{\text{Yuk}} = {\cal L}_{\text{Yuk, MCPM}} + {\cal L}_{\text{eff}}^{(2)} 
= {\cal L}_{\text{Yuk}}^{(1)} + {\cal L}_{\text{Yuk}}^{(2)}\;.
\end{equation}
Here we denote by  ${\cal L}_{\text{Yuk}}^{(i)}$, ($i=1,2$) the Yukawa terms connected to the Higgs fields~$\varphi_i$, respectively. 

We shall now write down~${\cal L}_{\text{Yuk}}^{(1)}$ explicitly,
\begin{equation} \label{A36}
\begin{split}
&{\cal L}_{\text{Yuk}}^{(1)} = 
- \begin{pmatrix} \bar{l}_{1 R}, &  \bar{l}_{2 R}, &  \bar{l}_{3 R} \end{pmatrix}
C_l^{(1)}
\begin{pmatrix} 
\varphi_1^\dagger \begin{pmatrix} \nu_{1L}\\ l_{1L} \end{pmatrix} \\
\varphi_1^\dagger \begin{pmatrix} \nu_{2L}\\ l_{2L} \end{pmatrix} \\
\varphi_1^\dagger \begin{pmatrix} \nu_{3L}\\ l_{3L} \end{pmatrix} \\
\end{pmatrix}\\
&
+ \begin{pmatrix} {\bar{u}}''_{1 R}, &  {\bar{u}}''_{2 R}, &  {\bar{u}}''_{3 R} \end{pmatrix}
\tilde{C}_u^{(1)}
\begin{pmatrix} 
\varphi_1^\trans \epsilon \begin{pmatrix} u'_{1L}\\ d_{1L} \end{pmatrix} \\
\varphi_1^\trans \epsilon \begin{pmatrix} u'_{2L}\\ d_{2L} \end{pmatrix} \\
\varphi_1^\trans \epsilon \begin{pmatrix} u'_{3L}\\ d_{3L} \end{pmatrix} 
\end{pmatrix}\\
&
- \begin{pmatrix} {\bar{d}}_{1 R}, &  {\bar{d}}_{2 R}, &  {\bar{d}}_{3 R} \end{pmatrix}
C_d^{(1)}
\begin{pmatrix} 
\varphi_1^\dagger \begin{pmatrix} u'_{1L}\\ d_{1L} \end{pmatrix} \\
\varphi_1^\dagger \begin{pmatrix} u'_{2L}\\ d_{2L} \end{pmatrix} \\
\varphi_1^\dagger \begin{pmatrix} u'_{3L}\\ d_{3L} \end{pmatrix} 
\end{pmatrix}\\
& + h.c.
\end{split}
\end{equation}
Here
\begin{equation} \label{A37}
C_l^{(1)} = 
\diag \begin{pmatrix} 
\delta_\chi b_{l\, 1}^{(1)},& 
\delta_\chi b_{l\, 2}^{(1)},& 
c_{l\, 3}^{(1)} + \delta_\chi b_{l\, 3}^{(1)}
\end{pmatrix},
\end{equation}
\begin{equation} \label{A37a}
\tilde{C}_u^{(1)} =
\begin{pmatrix} 
\delta_\chi {\tilde{b}}^{(1)}_{u\, 11} & \delta_\chi {\tilde{b}}^{(1)}_{u\, 12} & \delta_\chi {\tilde{b}}^{(1)}_{u\, 13} \\
\delta_\chi {\tilde{b}}^{(1)}_{u\, 21} & \delta_\chi {\tilde{b}}^{(1)}_{u\, 22} & \delta_\chi {\tilde{b}}^{(1)}_{u\, 23} \\
\delta_\chi {\tilde{b}}^{(1)}_{u\, 31} & \delta_\chi {\tilde{b}}^{(1)}_{u\, 32} & 
c_{u\, 3}^{(1)} + \delta_\chi {\tilde{b}}^{(1)}_{u\, 33} 
\end{pmatrix},
\end{equation}
\begin{equation} \label{A37b}
C_d^{(1)} = 
\diag \begin{pmatrix} 
\delta_\chi b_{d\, 1}^{(1)},& 
\delta_\chi b_{d\, 2}^{(1)},& 
c_{d\, 3}^{(1)} + \delta_\chi b_{d\, 3}^{(1)}
\end{pmatrix}.
\end{equation}
Now we study ${\tilde{C}_u}^{(1)}$. The product
\begin{equation} \label{A38}
{\tilde{C}_u}^{(1)} {\tilde{C}_u}^{(1) \dagger}
\end{equation}
is a hermitian, nonnegative ($3\times 3$) matrix. Therefore it can be diagonalized with a 
unitary matrix $U \in \mathit{U}(3)$,
\begin{equation} \label{A39}
U^\dagger {\tilde{C}_u}^{(1)} {\tilde{C}_u}^{(1) \dagger} U =
 \diag \begin{pmatrix} c_u^2, & c_c^2, & c_t^2 \end{pmatrix}.
\end{equation}
Here the diagonal elements of the matrix on the right-hand side of~\eqref{A39} must be nonnegative. We shall make the assumption that they are strictly positive,
\begin{equation} \label{A40}
c_u^2 >0, \qquad
c_c^2 >0, \qquad
c_t^2 >0\;.
\end{equation}
We get then from~\eqref{A39}
\begin{multline} \label{A41}
 \diag \begin{pmatrix} c_u^{-1}, & c_c^{-1}, & c_t^{-1} \end{pmatrix}\\
 \times
U^\dagger {\tilde{C}_u}^{(1)} {\tilde{C}_u}^{(1) \dagger} U
 \diag \begin{pmatrix} c_u^{-1}, & c_c^{-1}, & c_t^{-1} \end{pmatrix} = \unitmatrix_3\;,
 \end{multline}
 and we conclude that
\begin{equation} \label{A42}
 \diag \begin{pmatrix} c_u^{-1}, & c_c^{-1}, & c_t^{-1} \end{pmatrix}
 U^\dagger {\tilde{C}_u}^{(1)} = V
 \end{equation}
 satisfies
 \begin{equation} \label{A43}
 V^\dagger V = V V^\dagger = \unitmatrix_3\;.
 \end{equation}
 That is, $V \in \mathit{U}(3)$, and it will turn out to be the CKM matrix. 
 Now we define
 \begin{equation} \label{A44}
 C_u^{(1)} = 
 \diag \begin{pmatrix} c_u, & c_c, & c_t \end{pmatrix},
 \end{equation}
 and
 \begin{multline} \label{A45}
 \begin{pmatrix} 
 {\bar{u}_{1 R}}'(x),&
 {\bar{u}_{2 R}}'(x),&
 {\bar{u}_{3 R}}'(x),
 \end{pmatrix} 
 =\\
 \begin{pmatrix} 
 {\bar{u}_{1 R}}''(x),&
 {\bar{u}_{2 R}}''(x),&
 {\bar{u}_{3 R}}''(x)
 \end{pmatrix} U V\;.
 \end{multline}
 With this we get
 \begin{equation} \label{A46}
  \tilde{C}_u^{(1)} = U  C_u^{(1)}  V,
  \end{equation}
  and the Yukawa Lagrangian~\eqref{A36} reads
 \begin{equation} \label{A47}
\begin{split}
&{\cal L}_{\text{Yuk}}^{(1)} = 
- \begin{pmatrix} \bar{l}_{1 R}, &  \bar{l}_{2 R}, &  \bar{l}_{3 R} \end{pmatrix}
C_l^{(1)}
\begin{pmatrix} 
\varphi_1^\dagger \begin{pmatrix} \nu_{1L}\\ l_{1L} \end{pmatrix} \\
\varphi_1^\dagger \begin{pmatrix} \nu_{2L}\\ l_{2L} \end{pmatrix} \\
\varphi_1^\dagger \begin{pmatrix} \nu_{3L}\\ l_{3L} \end{pmatrix} \\
\end{pmatrix}\\
&
+ \begin{pmatrix} {\bar{u}}'_{1 R}, &  {\bar{u}}'_{2 R}, &  {\bar{u}}'_{3 R} \end{pmatrix}
V^\dagger C_u^{(1)} V
\begin{pmatrix} 
\varphi_1^\trans \epsilon \begin{pmatrix} u'_{1L}\\ d_{1L} \end{pmatrix} \\
\varphi_1^\trans \epsilon \begin{pmatrix} u'_{2L}\\ d_{2L} \end{pmatrix} \\
\varphi_1^\trans \epsilon \begin{pmatrix} u'_{3L}\\ d_{3L} \end{pmatrix} 
\end{pmatrix}\\
&
- \begin{pmatrix} {\bar{d}}_{1 R}, &  {\bar{d}}_{2 R}, &  {\bar{d}}_{3 R} \end{pmatrix}
C_d^{(1)}
\begin{pmatrix} 
\varphi_1^\dagger \begin{pmatrix} u'_{1L}\\ d_{1L} \end{pmatrix} \\
\varphi_1^\dagger \begin{pmatrix} u'_{2L}\\ d_{2L} \end{pmatrix} \\
\varphi_1^\dagger \begin{pmatrix} u'_{3L}\\ d_{3L} \end{pmatrix} 
\end{pmatrix}\\
& + h.c.
\end{split}
\end{equation}
 
 The $\mathit{SU}(2)_L$ doublet fields are 
 \begin{equation} \label{A48}
 \begin{pmatrix} u'_{ \alpha L}\\ d_{ \alpha L} \end{pmatrix}, \quad
 \begin{pmatrix} u_{ \alpha L}\\ d'_{ \alpha L} \end{pmatrix},  \quad \alpha = 1,2,3.
 \end{equation}
 The quark fields of definite mass are $u_\alpha$ and $d_\alpha$, where
 \begin{equation} \label{A49}
 u_\alpha = u_{ \alpha L} + u_{\alpha R}, \,
 d_\alpha = d_{ \alpha L} + d_{\alpha R}, \quad \alpha = 1,2,3,
 \end{equation}
 \begin{equation} \label{A50}
 \begin{pmatrix} u_{1 L}\\ u_{2 L}\\ u_{3 L} \end{pmatrix}
 = 
 V 
 \begin{pmatrix} u'_{1 L}\\ u'_{2 L}\\ u'_{3 L} \end{pmatrix},
 \begin{pmatrix} d_{1 L}\\ d_{2 L}\\ d_{3 L} \end{pmatrix}
 = 
 V^\dagger
 \begin{pmatrix} d'_{1 L}\\ d'_{2 L}\\ d'_{3 L} \end{pmatrix},
 \end{equation}
 \begin{equation} \label{A51}
 \begin{pmatrix} u_{1 R}\\ u_{2\,R}\\ u_{3 R} \end{pmatrix}
 = 
 V 
 \begin{pmatrix} u'_{1 R}\\ u'_{2 R}\\ u'_{3 R} \end{pmatrix},
 \begin{pmatrix} d_{1 R}\\ d_{2 R}\\ d_{3 R} \end{pmatrix}
 = 
 V^\dagger
 \begin{pmatrix} d'_{1 R}\\ d'_{2 R}\\ d'_{3 R} \end{pmatrix}.
 \end{equation}
 Indeed, with EWSB we have to replace~$\varphi_1(x)$ by~\eqref{eqA14} and we get
 \begin{multline} \label{A52}
 {\cal L}^{(1)}_{\text{Yuk}}(x) =
 -
 \left\{
 	\sum_l m_l\; \bar{l}(x) l(x) 
 	+
 	\sum_q m_q\; \bar{q}(x) q(x) 
 \right\}\\
 \times
 \left( 1 + \frac{\rho'(x)}{v_0} \right).
 \end{multline}
 Here the lepton and quark masses are obtained as follows from~\eqref{eqA14}, \eqref{A37}, \eqref{A37b}, and \eqref{A44},
 \begin{multline} \label{A53}
 m_e = \frac{v_0}{\sqrt{2}} \delta_\chi b_{l\, 1}^{(1)}, \quad
 m_\mu = \frac{v_0}{\sqrt{2}} \delta_\chi b_{l\, 2}^{(1)}, \\
 m_\tau = \frac{v_0}{\sqrt{2}} \left( c_{l\, 3}^{(1)} + \delta_\chi b_{l\, 3}^{(1)} \right),
 \end{multline}
 \begin{multline} \label{A54}
 m_d = \frac{v_0}{\sqrt{2}} \delta_\chi b_{d\, 1}^{(1)}, \quad
 m_s = \frac{v_0}{\sqrt{2}} \delta_\chi b_{d\, 2}^{(1)}, \\
 m_b = \frac{v_0}{\sqrt{2}} \left( c_{d\, 3}^{(1)} + \delta_\chi b_{d\, 3}^{(1)} \right),
 \end{multline}
 \begin{equation} \label{A55}
 m_u = \frac{v_0}{\sqrt{2}} c_u, \quad
 m_c = \frac{v_0}{\sqrt{2}} c_c, \quad
 m_t = \frac{v_0}{\sqrt{2}} c_t\;.
 \end{equation}
 
 So far we have followed closely the usual analysis of the Yukawa term of the SM Lagrangian;
 see for instance chapter~22.4 of~\cite{Nachtmann:1990ta}. Now we shall discuss the consequences of our ansatz here for the fermion masses and the CKM matrix.
 
 From~\eqref{A53} and \eqref{A54} we see that the masses of the $\tau$~lepton and the $b$ quark are (in general) slightly changed compared to the original values; see~\eqref{A26}. Remember that~$\delta_\chi$~\eqref{A19a} is a small parameter. In our numerical example~\eqref{eqA18} we have~$\delta_\chi = 7.3 \cdot 10^{-3}$. The masses of electron and muon are small compared to the $\tau$ mass because they are proportional to the small parameter $\delta_\chi$ and the same is true for the $d$- and $s$-quark masses compared to the $b$-quark mass.
 From~\eqref{14.3}, \eqref{A53}, and \eqref{A54} we find with our numerical 
 example~\eqref{eqA18} for~$\delta_\chi = 7.3 \cdot 10^{-3}$
\begin{alignat}{3} \label{A56}
&\frac{m_e}{m_\tau} = 3.95 \cdot 10^{-2} \delta_\chi, \qquad
&&\frac{m_\mu}{m_\tau} = 8.15  \delta_\chi, \nonumber \\
&\frac{m_d}{m_b} = 0.15 \delta_\chi, \qquad
&&\frac{m_s}{m_b} = 3.05 \delta_\chi. 
\end{alignat}
Now we consider the $u$-type-quark sector. 
With our ansatz we generate a hermitian matrix~(see~\eqref{A37a}, \eqref{A44}, \eqref{A46}, and \eqref{A47})
\begin{multline} \label{A57}
{\tilde{\tilde{C}}}_u^{(1)} = V^\dagger C_u^{(1)} V =
V^\dagger U^\dagger \tilde{C}_u^{(1)} =\\
\begin{pmatrix} 
\delta_\chi b^{(1)}_{u\, 11} & \delta_\chi b^{(1)}_{u\, 12} & \delta_\chi b^{(1)}_{u\, 13} \\
\delta_\chi b^{(1)}_{u\, 21} & \delta_\chi b^{(1)}_{u\, 22} & \delta_\chi b^{(1)}_{u\, 23} \\
\delta_\chi b^{(1)}_{u\, 31} & \delta_\chi b^{(1)}_{u\, 32} & 
c_{u\, 3}^{(1)} + \delta_\chi b^{(1)}_{u\, 33} 
\end{pmatrix}.
\end{multline}

It is easy to see that for small~$\delta_\chi$ the eigenvalues~\eqref{A44} of this matrix must be of the form
\begin{equation} \label{A58}
c_u = \delta_\chi b_{u\, 1}^{(1)}, \quad
c_c = \delta_\chi b_{u\, 2}^{(1)}, \quad
c_t = c_{u\, 3}^{(1)} + \delta_\chi b_{u\, 3}^{(1)}.
\end{equation}
Indeed, we can first diagonalize the 1--2 sector of the matrix~$\tilde{\tilde{C}}_u^{(1)}$ of~\eqref{A57}. We get then a matrix~${\hat{C}_u}^{(1)}$. Next we use standard perturbation theory for the eigenvalues of hermitian matrices. Here we choose as perturbing term the matrix formed by the 13, 23, 31, and 32 components of~~${\hat{C}_u}^{(1)}$.
Therefore, we get for the $u$-, $c$-, and $t$-quark masses in analogy to~\eqref{A53} and \eqref{A54},
\begin{multline} \label{A59}
 m_u = \frac{v_0}{\sqrt{2}} \delta_\chi b_{u\, 1}^{(1)}, \quad
 m_c = \frac{v_0}{\sqrt{2}} \delta_\chi b_{u\, 2}^{(1)}, \\
 m_t = \frac{v_0}{\sqrt{2}}  c_t =
 \frac{v_0}{\sqrt{2}} 
 \left( c_{u\, 3}^{(1)} + \delta_\chi b_{u\, 3}^{(1)} \right).
 \end{multline}
The $u$ and $c$ masses are small compared to the $t$-quark mass since they are proportional to~$\delta_\chi$. Experimentally we have $c_t = 0.99$; see~\eqref{A26}. 
For our numerical example~\eqref{eqA18}, \eqref{A19a}, and the mass ratios from~\eqref{14.3} we get
\begin{multline} \label{A60}
\frac{m_u}{m_t} = \delta_\chi 1.7 \cdot 10^{-3}, \\
\frac{m_c}{m_t} = \delta_\chi \frac{b_{u\, 2}^{(1)}\; v_0/\sqrt{2} }{c_t\; v_0/\sqrt{2} }
= \delta_\chi \frac{b_{u\, 2}^{(1)}}{ c_t}
=\delta_\chi 1.01\;.
\end{multline}
Remember that our example was chosen to have~$b_{u\, 2}^{(1)} \equiv b^{(c)}= 1$; see~\eqref{eqA17}. The top-quark mass in~\eqref{A59} is slightly changed compared to the original mass in~\eqref{A26}. 

Next we shall discuss what our result~\eqref{A57} for ${\tilde{\tilde{C}}}^{(1)}_u = V^\dagger C_u^{(1)} V$ implies for the matrix~$V$ which relates the $\mathit{SU}(2)_L$ doublet partners $u'_{\alpha L}$ of the fields~$d_{\alpha L}$ to the fields $u_\alpha$ of definite mass; see~\eqref{A48}--\eqref{A52}. Thus, $V$ is the usual CKM matrix. We find it convenient to write down explicitly ${\tilde{\tilde{C}}}^{(1)}_{u}=V^\dagger C_u^{(1)} V$, $V^\dagger V$, and $V V^\dagger$; see~\eqref{A61}--\eqref{A63}. 

\begin{widetext}
\begin{equation} \label{A61}
\tilde{\tilde{C}}_u^{(1)} =
V^\dagger C_u^{(1)} V=
\begin{pmatrix}
c_u V^*_{11} V_{11} + c_c V^*_{21} V_{21}+ c_t V^*_{31} V_{31} &
c_u V^*_{11} V_{12}+ c_c V^*_{21} V_{22}+ c_t V^*_{31} V_{32} &
c_u V^*_{11} V_{13}+ c_c V^*_{21} V_{23}+ c_t V^*_{31} V_{33}\\
c_u V^*_{12} V_{11}+ c_c V^*_{22} V_{21}+ c_t V^*_{32} V_{31} &
c_u V^*_{12} V_{12}+ c_c V^*_{22} V_{22}+ c_t V^*_{32} V_{32} &
c_u V^*_{12} V_{13}+ c_c V^*_{22} V_{23}+ c_t V^*_{32} V_{33}\\
c_u V^*_{13} V_{11}+ c_c V^*_{23} V_{21}+ c_t V^*_{33} V_{31} &
c_u V^*_{13} V_{12}+ c_c V^*_{23} V_{22}+ c_t V^*_{33} V_{32} &
c_u V^*_{13} V_{13}+ c_c V^*_{23} V_{23}+ c_t V^*_{33} V_{33}
\end{pmatrix}
\end{equation}
\begin{equation} \label{A62}
V^\dagger V =
\begin{pmatrix}
V^*_{11} V_{11}+V^*_{21} V_{21}+V^*_{31} V_{31} &
V^*_{11} V_{12}+V^*_{21} V_{22}+V^*_{31} V_{32} &
V^*_{11} V_{13}+V^*_{21} V_{23}+V^*_{31} V_{33}\\
V^*_{12} V_{11}+V^*_{22} V_{21}+V^*_{32} V_{31} &
V^*_{12} V_{12}+V^*_{22} V_{22}+V^*_{32} V_{32} &
V^*_{12} V_{13}+V^*_{22} V_{23}+V^*_{32} V_{33}\\
V^*_{13} V_{11}+V^*_{23} V_{21}+V^*_{33} V_{31} &
V^*_{13} V_{12}+V^*_{23} V_{22}+V^*_{33} V_{32} &
V^*_{13} V_{13}+V^*_{23} V_{23}+V^*_{33} V_{33}
\end{pmatrix}
= \unitmatrix_3\;,
\end{equation}

\begin{equation} \label{A63}
V V^\dagger =
\begin{pmatrix}
V_{11} V^*_{11}+V_{12} V^*_{12}+V_{13} V^*_{13} &
V_{11} V^*_{21}+V_{12} V^*_{22}+V_{13} V^*_{23} &
V_{11} V^*_{31}+V_{12} V^*_{32}+V_{13} V^*_{33}\\
V_{21} V^*_{11}+V_{22} V^*_{12}+V_{23} V^*_{13} &
V_{21} V^*_{21}+V_{22} V^*_{22}+V_{23} V^*_{23} &
V_{21} V^*_{31}+V_{22} V^*_{32}+V_{23} V^*_{33}\\
V_{31} V^*_{11}+V_{32} V^*_{12}+V_{33} V^*_{13} &
V_{31} V^*_{21}+V_{32} V^*_{22}+V_{33} V^*_{23} &
V_{31} V^*_{31}+V_{32} V^*_{32}+V_{33} V^*_{33}
\end{pmatrix}
= \unitmatrix_3\;,
\end{equation}
\end{widetext}

From~\eqref{A57} and \eqref{A61} we see that 
\begin{equation} \label{A63a}
{\tilde{\tilde{C}}}^{(1)}_{u\, 13}, \quad
{\tilde{\tilde{C}}}^{(1)}_{u\, 23}, \quad
{\tilde{\tilde{C}}}^{(1)}_{u\, 31}, \quad
{\tilde{\tilde{C}}}^{(1)}_{u\, 32}
= {\cal O} (\delta_\chi)\;.
\end{equation}
We have already~$c_u$ and $c_c$ of order~$\delta_\chi$. Therefore, with~$c_t \approx 1$ we must have 
\begin{equation} \label{A64}
V^*_{31} V_{33}, \quad
V^*_{32} V_{33}, \quad
V^*_{33} V_{31}, \quad
V^*_{33} V_{32}
= {\cal O} (\delta_\chi)\;.
\end{equation}
It then follows from~\eqref{A63} that
\begin{equation} \label{A65}
\begin{split}
&|V_{31}|^2 + |V_{32}|^2 + |V_{33}|^2 =1,\\
&\left( |V_{31}|^2 + |V_{32}|^2 \right) |V_{33}|^2=
|V_{33}|^2 - |V_{33}|^4 = {\cal O} (\delta_\chi^2)\;,
\end{split}
\end{equation}

This implies that either $|V_{33}|^2$ or $1-|V_{33}|^2$ must be of order~$\delta_\chi^2$. Since we have $V_{33} \to 1$ for $\delta_\chi \to 0$ we conclude that
\begin{equation} \label{A66}
|V_{33}|^2 = 1 - {\cal O}(\delta_\chi^2), \quad
|V_{33}| = 1 - {\cal O}(\delta_\chi^2).
\end{equation}
From~\eqref{A62} we get then 
\begin{equation} \label{A67}
|V_{13}|^2 + |V_{23}|^2 =
1- |V_{33}|^2 = {\cal O} (\delta_\chi^2)\;,
\end{equation}
which implies
\begin{equation} \label{A68}
 |V_{13}|, \quad |V_{23}| = {\cal O} (\delta_\chi).
\end{equation}

To summarize: from our ansatz of generating the masses of the fermions of the first and second family as well as the CKM matrix~$V$ from interactions at a high mass scale~$m_\chi \approx 10$~TeV we find that the CKM matrix must have $|V_{33}|$ very close to 1, and $|V_{13}|$, $|V_{23}|$, $|V_{31}|$, $|V_{32}|$ small, of order~$\delta_\chi$. We see from~\eqref{A61} that the 1--2 sector of $\tilde{\tilde{C}}^{(1)}_u$ is then already of order~$\delta_\chi$ since $c_u$ and $c_c$ are of order~$\delta_\chi$. Thus, 
$|V_{11}|$, $|V_{12}|$, $|V_{21}|$, and $|V_{22}|$ can be of ``normal'' size, that is, of order~1. All this checks rather nicely with the experimental results for the absolute values of~$|V_{ij}|$; see~Eq.~(12.27) of~\cite{ParticleDataGroup:2024cfk}. We find from the central values quoted there and with~$\delta_\chi = 7.3 \cdot 10^{-3}$ from~\eqref{eqA18}, \eqref{A19a},
\begin{equation} \label{A69}
\begin{split}
&|V_{13}| = 0.003732 = 0.511 \;\delta_\chi,\\
&|V_{23}| = 0.04183 = 5.73 \;\delta_\chi,\\
&|V_{31}| = 0.00858 = 1.18 \;\delta_\chi,\\
&|V_{32}| = 0.04111= 5.63 \;\delta_\chi,
\end{split}
\end{equation}
\begin{equation} \label{A70}
1- |V_{33}| = 8.82 \cdot 10^{-4} =
16.55 \cdot \delta_\chi^2.
\end{equation}
On the other hand $|V_{12}|$ and $|V_{21}|$ are around 0.22, that is, indeed of order one. 

The next task is to discuss the coupling of the Higgs field~$\varphi_2(x)$ to the fermions. The MCPM part can be read off from~\eqref{A25}. Now we make an ansatz for the current~$J_2^{(f)}(x)$ in~\eqref{A28a}. We set
\begin{multline} \label{A71}
J_2^{(f)}(x) = b_\mu^{(2)} \bar{\mu}_R(x) \begin{pmatrix} \nu_{\mu L}(x)\\ \mu_L(x) \end{pmatrix}
\\ 
-
b_c^{(2)} \epsilon \begin{pmatrix} \bar{c}_L(x)\\ s'_L(x) \end{pmatrix} c_R(x) 
+
b_s^{(2)} \bar{s}_R(x)  \begin{pmatrix} c'_L(x)\\ s_L(x) \end{pmatrix}.
\end{multline}
Here we have from~\eqref{A48}--\eqref{A51}
\begin{equation} \label{A72}
c'_L(x) \equiv u'_{2 L}(x) = \sum_{\alpha=1}^3 V_{2\, \alpha}^\dagger u_{\alpha L}(x),
\end{equation}
and similarly
\begin{equation} \label{A73}
s'_L(x) \equiv d'_{2 L}(x) = \sum_{\alpha=1}^3 V_{2\, \alpha} d_{\alpha L}(x).
\end{equation}
We emphasize that~\eqref{A71} is a simple ansatz. Certainly we could extend this ansatz producing then additional small coupling terms of order~$\delta_\chi$ in the Yukawa Lagrangian~${\cal L}^{(2)}_{\text{Yuk}}$ below. 
Inserting the current~\eqref{A71} in ${\cal L}_{\text{eff}}^{(2,2)}$ from~\eqref{A33} and adding the $\varphi_2$ part of the MCPM Yukawa interaction from~\eqref{A25} we get the following Yukawa term 
${\cal L}^{(2)}_{\text{Yuk}}$, \eqref{A35},
\begin{equation} \label{A74}
\begin{split}
{\cal L}^{(2)}_{\text{Yuk}} = & (c_\tau^{(1)} + \delta_\chi b_\mu^{(2)}) \bar{\mu}_R \varphi_2^\dagger
\begin{pmatrix} \nu_{\mu L}\\ \mu_L \end{pmatrix}
\\ & 
-
(c_t^{(1)} + \delta_\chi b_c^{(2)}) \bar{c}_R \varphi_2^\trans \epsilon
\begin{pmatrix} c_L\\ s'_L \end{pmatrix}
\\ & 
+
(c_b^{(1)} + \delta_\chi b_s^{(2)}) \bar{s}_R \varphi_2^\dagger
\begin{pmatrix} c'_L\\ s_L \end{pmatrix} + h.c.
\end{split}
\end{equation}
Now we set
\begin{equation} \label{A75}
\varphi_2(x) = \begin{pmatrix} H^+(x)\\ \frac{1}{\sqrt{2}}(h'(x) + i h''(x)) \end{pmatrix}
\end{equation}
from~(38) of \cite{Maniatis:2007de} and assume for simplicity
\begin{equation} \label{A76}
\begin{split}
&c_\tau^{(1)} + \delta_\chi b_\mu^{(2)} = c_\tau = \frac{\sqrt{2}}{v_0} m_\tau, \\
&c_t^{(1)} + \delta_\chi b_c^{(2)} = c_t = \frac{\sqrt{2}}{v_0} m_t, \\
&c_b^{(1)} + \delta_\chi b_s^{(2)} = c_b = \frac{\sqrt{2}}{v_0} m_b.
\end{split}
\end{equation}
We get then with~\eqref{A72} and \eqref{A73}
\begin{align}\label{A77}
& {\cal L}_{\mathrm{Yuk}}^{(2)} =
h' \left[ \frac{m_\tau}{v_0} \bar{\mu}\mu +   \frac{m_t}{v_0} \bar{c}c  + \frac{m_b}{v_0} \bar{s}s \right]
\notag\\ &
+
h'' \left[ i \frac{m_\tau}{v_0} \bar{\mu}\gamma_5 \mu - i  \frac{m_t}{v_0} \bar{c} \gamma_5 c + i \frac{m_b}{v_0} \bar{s}\gamma_5 s \right]
\notag\\ &
+
H^+ \Bigg[ \frac{m_\tau}{\sqrt{2}v_0} \bar{\nu}_\mu (1+\gamma_5) \mu 
\notag\\ & \quad
	-   \frac{m_t}{\sqrt{2} v_0} \bar{c} (1-\gamma_5)
	\left( V_{21} d + V_{22} s + V_{23} b \right)
\notag\\ & \quad
	+   \frac{m_b}{\sqrt{2} v_0} 
	\left( \bar{u} V_{12}  + \bar{c} V_{22} + \bar{t} V_{32} \right) (1+\gamma_5) s
	\Bigg]
\notag\\ &
+
H^- \Bigg[ \frac{m_\tau}{\sqrt{2}v_0} \bar{\mu} (1-\gamma_5) \nu_\mu
\notag\\ & \quad
	-   \frac{m_t}{\sqrt{2} v_0} 
	\left( V_{21}^* \bar{d} + V_{22}^* \bar{s} + V_{23}^* \bar{b} \right) (1+\gamma_5) c
\notag\\ & \quad
	+   \frac{m_b}{\sqrt{2} v_0} \bar{s} (1-\gamma_5)
	\left( V_{12}^* u  + V_{22}^* c + V_{32}^* t \right) 
	\Bigg].
\end{align}
The Lagrangian~\eqref{A77} is a natural generalisation of  the MCPM Yukawa part connected with~$\varphi_2$ in~\eqref{A25}. There is no claim that it is the most general ansatz. But we choose~\eqref{A77} as part of the definition of our model MCPM'. We summarize the main features of this model, as we obtained them in this Appendix~\ref{appA}, in the conclusion section.

 %%%%%%%%%%%%%%%%%%%%%%%%%%%%%%%%%%%%% Appendix B
%%%%%%%%%%%%%%%%%%%%%%%%%%%%%%%%%%%%
\section{The cross section $\sigma(p(p_1)+p(p_2) \to \gamma \gamma + X)$}
\label{appB}

Here we calculate the cross section $\sigma(p(p_1)+p(p_2) \to \gamma \gamma + X)$ taking the finite width of the Higgs boson~$h''$, $\Gamma_{h''}$, into account. 
We consider the variation of the cross section for $h''$ production with the actual $h''$ mass, that is, with $m_{\gamma\gamma}$, and similarly the variation of $\Gamma(h''\to \gamma \gamma)$. This gives the following formula:
\begin{multline} \label{B1}
\sigma(p(p_1)+p(p_2) \to \gamma + \gamma + X) =
\\
\left[ B(h'' \to \gamma \gamma) \times
\sigma(p(p_1)+p(p_2) \to h'' + X) \right]_{\text{eff}} =
\\ 
\int
\frac{ d \sigma}{d m^2_{\gamma \gamma}}
\big( p(p_1)+p(p_2) \to \gamma(k_1) + \gamma(k_2) + X \big)_{k^2=(k_1+k_2)^2=m_{\gamma \gamma}^2}
\\
\times
 2 m_{\gamma \gamma} d m_{\gamma \gamma} \;,
\end{multline}
with
\begin{multline} \label{B2}
\frac{ d \sigma}{d m^2_{\gamma \gamma}} 
\big( p(p_1)+p(p_2) \to \gamma(k_1) + \gamma(k_2) + X \big)_{k^2 =m_{\gamma \gamma}^2} =
\\
\left.B(h''(k) \to \gamma \gamma)\right|_{k^2=m_{\gamma \gamma}^2}
\\
\times
\left.\sigma(p(p_1)+p(p_2) \to h''(k) + X)\right|_{k^2=m_{\gamma \gamma}^2}
\\
\times
\frac{m_{\gamma \gamma}}{m_{h''}}
\frac{\left.\Gamma_{h''}\right|_{k^2=m_{\gamma \gamma}^2}}{\Gamma_{h''}}
\frac{m_{h''} \Gamma_{h''}/\pi}{(m^2_{\gamma \gamma} - m^2_{h''})^2 + m^2_{h''} \Gamma^2_{h''}} \;.
\end{multline}
Here, $m_{h''}$ and $\Gamma_{h''}$ denote the nominal mass and the total width of the $h''$ boson.
The width $\left.\Gamma_{h''}\right|_{k^2=m_{\gamma \gamma}^2}$ denotes the total width with the mass of $m_{h''}$ set to the invariant mass $m_{\gamma \gamma}$ of the two photons. In the integral in $\eqref{B1}$ we vary the integration limits with a parameter~$\kappa$:
\begin{equation} \label{B3}
m_{h''} - \kappa \Gamma_{h''} \le m_{\gamma \gamma} \le m_{h''} + \kappa \Gamma_{h''}, 
\quad \text{with } \kappa = 1/2, 1, 2\;.
\end{equation}
We find for the total cross section~\eqref{B1} depending on~$\kappa$:
\begin{align} \label{B4}
& \kappa = 1/2:  \; && \sigma = 0.0051~\text{pb}, \nonumber\\
& \kappa = 1:  \; && \sigma = 0.0072~\text{pb},\\
& \kappa = 2:  \; && \sigma = 0.0085~\text{pb} \nonumber.
\end{align}
 As expected we see that the total cross section $\sigma(p(p_1)+p(p_2) \to \gamma + \gamma + X)$ approaches, for a large interval corresponding to a large~$\kappa$, the total cross section of 0.01~pb from~\eqref{eq4.1}. For $\kappa=1/2$, that is, for an integration interval of total length~$\Gamma_{h''}$ around the nominal 
 invariant mass $m_{h''}$, the cross section drops to about half this value. 
 
 We emphasize that once good data are available one will have to perform a theoretical calculation taking not only the finite width but also higher order effects into account. This will presumably change our results~\eqref{B4} somewhat.

% =============================================================================
%\bibliographystyle{JHEP}
\bibliography{references}

\end{document}